\documentstyle[12pt,fleqn,elsevier,epsfig,Here]{article}
\begin{document}           
\begin{center}
{\bf INSTITUT~F\"{U}R~KERNPHYSIK,~UNIVERSIT\"{A}T~FRANKFURT}\\
60486 Frankfurt, August--Euler--Strasse 6, Germany
\end{center}

\hfill IKF--HENPG/6--94
\begin{center}
\end{center}
\vspace{.5cm}

\begin{center}
\begin{Large}
{\bf Charged Particle Production in Proton-, Deuteron-, Oxygen- and Sulphur-Nucleus
Collisions at 200 GeV per Nucleon}
\end{Large}
\end{center}


{\bf Abstract}

The transverse momentum and rapidity distributions of net
protons and negatively charged hadrons have been measured for
minimum bias proton--nucleus and deuteron--gold interactions, as well as 
central oxygen--gold and 
sulphur--nucleus collisions at 200 GeV per nucleon.
The rapidity density of net protons at midrapidity in central
nucleus--nucleus collisions increases both with target mass for sulphur 
projectiles and with the projectile mass for a gold target.
The shape of the rapidity distributions of net protons forward of
midrapidity for d+Au and central S+Au collisions is similar. 
The average rapidity loss is larger than 2 units of rapidity for
reactions with the gold target.	
The transverse momentum spectra of net protons for all reactions 
can be described
by a thermal distribution with `temperatures' between $145 \pm 11$ MeV 
(p+S interactions) and $244 \pm 43$ MeV (central S+Au collisions).
The multiplicity of negatively charged hadrons increases with the
mass of the colliding system.  
The shape of the transverse momentum spectra of negatively
charged hadrons changes from minimum bias p+p and p+S interactions to 
p+Au and central nucleus-nucleus collisions. The mean transverse
momentum is almost constant in the vicinity of midrapidity and
shows little variation with the target and projectile masses.
The average number of produced negatively charged hadrons per participant
baryon increases slightly from p+p, p+A to central S+S,Ag collisions. 

\vspace{5cm}

\newpage

\begin{center}
{\bf
 The NA35 Collaboration:}
\end{center}
\vspace{0.5cm}
\noindent
T.~Alber$^{10}$, H.~Appelsh\"auser$^{6}$, J.~B\"achler$^{5,a}$, 
J.~Bartke$^4$, H.~Bia{\l}kowska$^{12}$, M.A.~Bloomer$^{3}$,
R.~Bock$^5$, W.J.~Braithwaite$^{10}$, D.~Brinkmann$^{6}$,
R.~Brockmann$^5$, P.~Bun\v{c}i\'c$^{5,b}$, P.~Chan$^{10}$, 
J.G.~Cramer$^{10}$, P.B.~Cramer$^{10}$, I.~Derado$^{9}$, 
V.~Eckardt$^{9}$, J.~Eschke$^{6,c}$, C.~Favuzzi$^2$, D.~Ferenc$^{13}$,
B.~Fleischmann$^5$, P.~Foka$^{5,b}$, P.~Freund$^{9}$, M.~Fuchs$^6$,
M.~Ga\'zdzicki$^{6}$, E.~G{\l}adysz$^{4}$, J.~Grebieszkow$^{11}$,
J.~G\"unther$^{6}$,
J.W.~Harris$^{3,d}$, M.~Hoffmann$^7$,
P.~Jacobs$^3$, S.~Kabana$^{6,e}$, K.~Kadija$^{9,13}$, R.~Keidel$^8$,  
M.~Kowalski$^{4}$, A.~K\"uhmichel$^6$,
J.Y.~Lee$^6$, A.~Ljubi\v{c}i\'c~jr.$^{13,f}$,
S.~Margetis$^{3,g}$, J.T.~Mitchell$^{3,f}$, R.~Morse$^3$, 
E.~Nappi$^2$, G.~Odyniec$^3$, G.~Pai\'c$^{5,13}$,
A.D.~Panagiotou$^1$, A.~Petri\-dis$^1$, A.~Piper$^8$,
F.~Posa$^2$, A.M.~Poskanzer$^3$,
F.~P\"uhlhofer$^8$, 
W.~Rauch$^{9}$, R.~Renfordt$^6$, W.~Retyk$^{11}$, D.~R\"ohrich$^{6}$, 
G.~Roland$^6$, H.~Rothard$^6$, K.~Runge$^{7}$,
A.~Sandoval$^5$, N.~Schmitz$^{9}$,
E.~Schmoetten$^{7}$, R.~Sendelbach$^{6}$,
P.~Seyboth$^{9}$, J.~Seyerlein$^{9}$,
E.~Skrzypczak$^{11}$, P.~Spinelli$^2$,  
R.~Stock$^6$, H.~Str\"obele$^6$,
L.~Teitel\-baum$^3$, S.~Tonse$^3$, 
T.A.~Trainor$^{10}$, G.~Vasileiadis$^1$, M.~Vassiliou$^1$,
G.~Vesztergombi$^{9}$, D.~Vranic$^{13}$, S.~Wenig$^6$, B.~Wosiek$^{9,4}$,
X.~Zhu$^{10}$

\vspace{0.3cm}
\noindent
$^1$Department of Physics, University~of Athens, Athens, Greece,  
$^2$Dipartimento di Fisica, Universit\`a~di Bari and INFN Bari, Bari, 
Italy, 
$^3$Lawrence Berkeley Laboratory, Berkeley, CA, USA, 
$^4$Institute~of Nuclear Physics, Cracow, Poland, 
$^5$Gesellschaft f\"ur Schwerionenforschung (GSI),
Darmstadt, Fed. Rep. of Germany,  
$^6$Fachbereich Physik der Universit\"at, Frankfurt,
Fed. Rep. of Germany (IKF), 
$^7$Fakult\"at f\"ur Physik der Universit\"at, Freiburg,
Fed. Rep. of Germany, 
$^8$Fachbereich Physik der Universit\"at, Marburg,
Fed. Rep. of Germany, 
$^{9}$Max--Planck--Institut f\"ur Physik, M\"unchen,
Fed. Rep. of Germany, 
$^{10}$University of Washington, Seattle, USA, 
$^{11}$Institute~for Experimental Physics, University~of Warsaw, 
Warsaw, Poland,  
$^{12}$Institute~for Nuclear Studies, Warsaw, Poland, 
$^{13}$Rudjer Bo\v{s}kovi\'c Institute, Zagreb. \\
\noindent
$^{a}$Now at CERN\\
$^{b}$Now at IKF\\
$^{c}$Now at GSI\\
$^{d}$Now at Yale University, New Haven, CT, USA\\ 
$^{e}$Now at University of Bern, Bern, Switzerland\\ 
$^{f}$Now at Brookhaven National Laboratory, Upton, NY, USA\\ 
$^{g}$Now at Kent State University, Kent, OH, USA

\newpage

\section{Introduction}

In order to investigate and
understand the dynamics of relativistic nucleus-nucleus
collisions, it is important to have information
on the
rapidity and
transverse momentum distributions of
nucleons participating in the interaction 
and of particles produced in the
collisions, all as a function of the size of the system and of the impact parameter.
The differences
of particle distributions
from proton--proton, proton--nucleus and finally 
nucleus--nucleus interactions may provide the key to understanding
any non-hadronic effects that may appear in central nucleus--nucleus 
collisions. Ultimately these distributions might be used to
determine the behavior of strongly interacting matter at high 
densities and temperatures.

New results from the measurement of collisions of very heavy nuclei
at the BNL AGS and CERN SPS \cite{QM95,QM96} can best be understood
by comparison to simpler systems of lighter projectiles and
targets.
%
%
We report results from the
NA35 experiment at the CERN SPS on
the rapidity and transverse momentum
distributions of negatively charged hadrons $h^-$ and participant protons,
measured in proton--nucleus, deuteron--gold,
oxygen--gold and sulphur--nucleus collisions at 200 GeV per nucleon.
A small subset of the present data (S+S collisions)
has already been published \cite{Bae94} 
and has been compared to various theoretical predictions
(\cite{Sorge,Ornik,Soll}).
The transverse momentum and rapidity
distributions provide information on
the degree
of stopping, thermalization, expansion and flow effects
in the collisions.
Furthermore, the participant proton rapidity distributions
may be used to
determine the baryon density or the baryo-chemical potential as a 
function of rapidity, which constrains
model calculations of the dynamics of particle production.
The present study is complementary to
transverse and forward energy measurements
already reported from NA35 \cite{Bae91}
which provided global information on the collision dynamics. It is also
supplementary to previous
strange particle production measurements \cite{Bam89,Bam90,Bae93,Alb94} 
which are the basis for the determination of the chemical freeze-out conditions.
All of this information 
helps to obtain a consistent
picture of the entire
reaction process. In particular, it will facilitate 
detailed comparisons with model calculations 
(e.g. \cite{Sorge,Venus,Cley,Stachel,Leon}).

\newpage

\section{Experiment NA35}

Experiment NA35 at the CERN SPS studies collisions of p, d,
$^{16}$O and $^{32}$S projectiles of 200 GeV per nucleon incident
energy ($\sqrt{s_{NN}}$=19.4 GeV) with nuclear targets. 
The main detectors are two large-volume tracking devices: 
a streamer chamber (SC) inside a 1.5 Tesla vertex magnet (VTM) and a 
time projection chamber (TPC) positioned downstream of the magnet. 
Both detectors 
record the space trajectories of charged particles from which 
the particle momenta are derived. The acceptances of the detectors 
for negatively charged hadrons are complementary to each other and
cover the full phase space with some overlap. Central
events are usually selected by the absence of projectile spectators
in the veto calorimeter placed in the beam line downstream  
of all detectors (see e.g. \cite{Bae94,Bae91,Alb94,Bri95}).

The objective of the experiment is to measure the hadronic final
state in nucleus--nucleus collisions. 
In addition to negatively and positively
charged particles, weak decays of charged and neutral particles  
can be detected in the streamer chamber. 
The analysis of the streamer chamber pictures has been fully
automated \cite{Bri95}, which not only resulted in an increase of the
measuring rate by an order
of magnitude but also allowed for a complete simulation of the
detector based on the GEANT-package. 
The results discussed in the following are mainly based on the 
automated analysis 
but traditional measurements have been included where appropriate.
The TPC identifies 
charged particles by
multiple sampling of the specific ionisation with a
$dE/dx$ resolution 
of about 5-6$\%$. It covers the momentum range of the relativistic rise of the energy loss. 
This makes particle identification using 
statistical methods possible.
Combining both detectors, spectra of negatively charged hadrons $h^-$, 
$K^0_S$, $K^+$, $K^-$, net protons ($p-\overline{p}$), $\Lambda$, 
$\overline{\Lambda}$ and $\overline{p}$ are obtained.
Thus information on practically all baryons and 
mesons in the final hadronic state can be accessed
with large phase space coverage. 
This paper will focus on rapidity
and transverse momentum spectra of net (mainly participating) protons and 
negatively charged hadrons (primarily pions).

In order to accomplish large phase space coverage in collisions
with very different particle multiplicities (from p+S to S+Au) and 
track densities in the laboratory, data were taken in three different 
experimental setups. 
Proton- and the oxygen-induced
reactions were studied using the (standard)
streamer chamber configuration (SETUP I) described in \cite{Bam89,Bri95,SCNIM}.
The nuclear target was placed 11 cm upstream of the 
2.0 $\times$ 1.2 $\times$ 0.72 m$^3$ 
streamer chamber (external target) which is the main tracking device inside the 1.5 T 
magnetic field of the superconducting vertex magnet. 
In addition p+S interactions were recorded with a target placed inside
the streamer chamber (internal target). 
Charged particles
created tracks of streamers of about 1 mm diameter and 1 cm length 
along their path 
through the sensitive volume. Three cameras, each equipped with a two-stage
magnetically focusing image intensifier of about 2500-fold gain,
recorded the events with a demagnification of 30 on 70 mm film. 
For the $^{32}$S induced reactions the track density in the central
region
along the beam direction ($\Theta_{lab} \leq 5^{\circ}$) is so high that
individual tracks cannot be
separated due to the projective readout of the streamer chamber by
stereoscopic cameras. Therefore, the streamer chamber geometry was modified by
the insertion of an opaque foil of a trapezoidal shape close to the
beam axis 
reducing the projected track density in that region.
In addition, another opaque foil screened the lower half 
of the streamer chamber
in order to further reduce 
the track density. This modified configuration is called SETUP II.

\begin{figure}[hbt]
\centerline{\epsfig{file=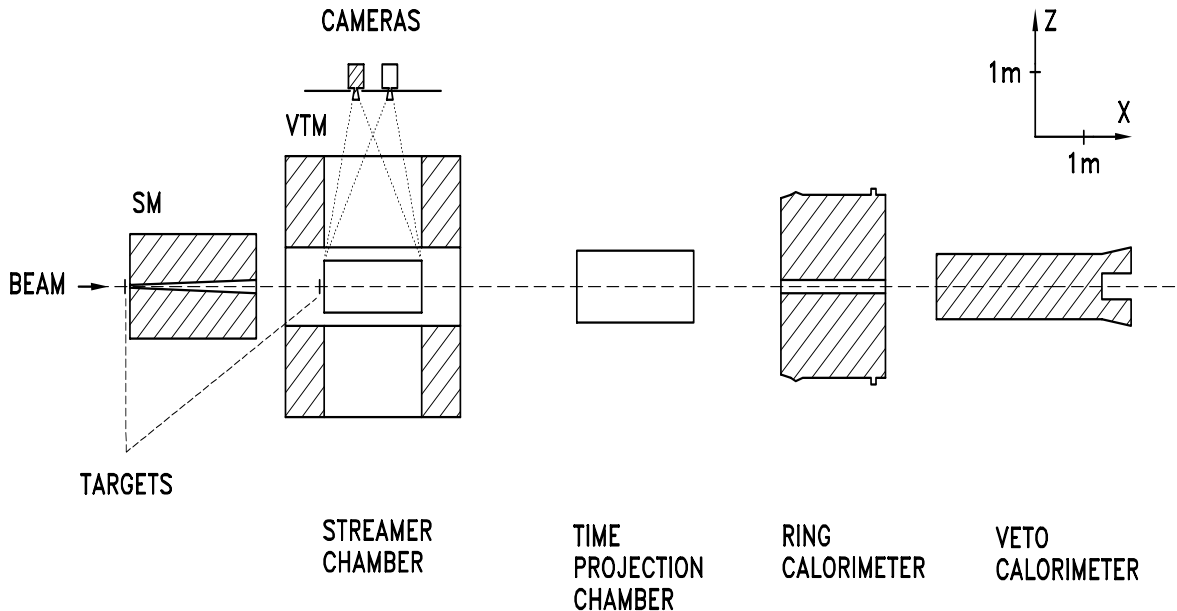,height=8.0cm,width=15.5cm,angle=90.}}
\caption{Experimental setup of experiment NA35 at the
CERN SPS. The target is placed either in front of the streamer 
chamber (SETUP I and II) or in front of the sweeper magnet SM (SETUP III).}
\label{sweeper}
\end{figure}

\begin{figure}[hbt]
\begin{minipage}[t]{8cm}
\centerline{\epsfig{file=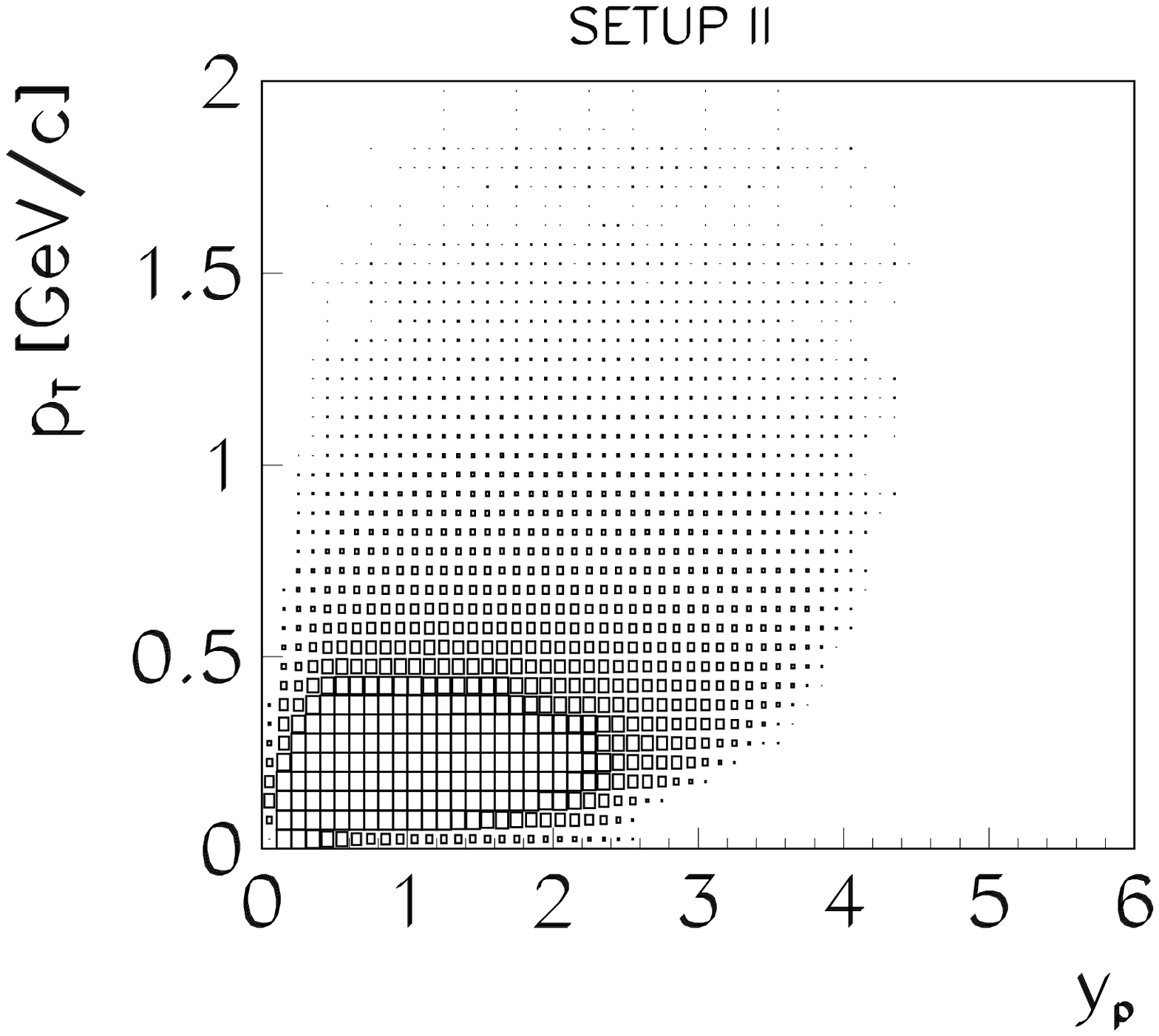,height=8.0cm,width=8cm}}
\end{minipage}\hspace*{0.6cm}
\begin{minipage}[t]{8cm}
\centerline{\epsfig{file=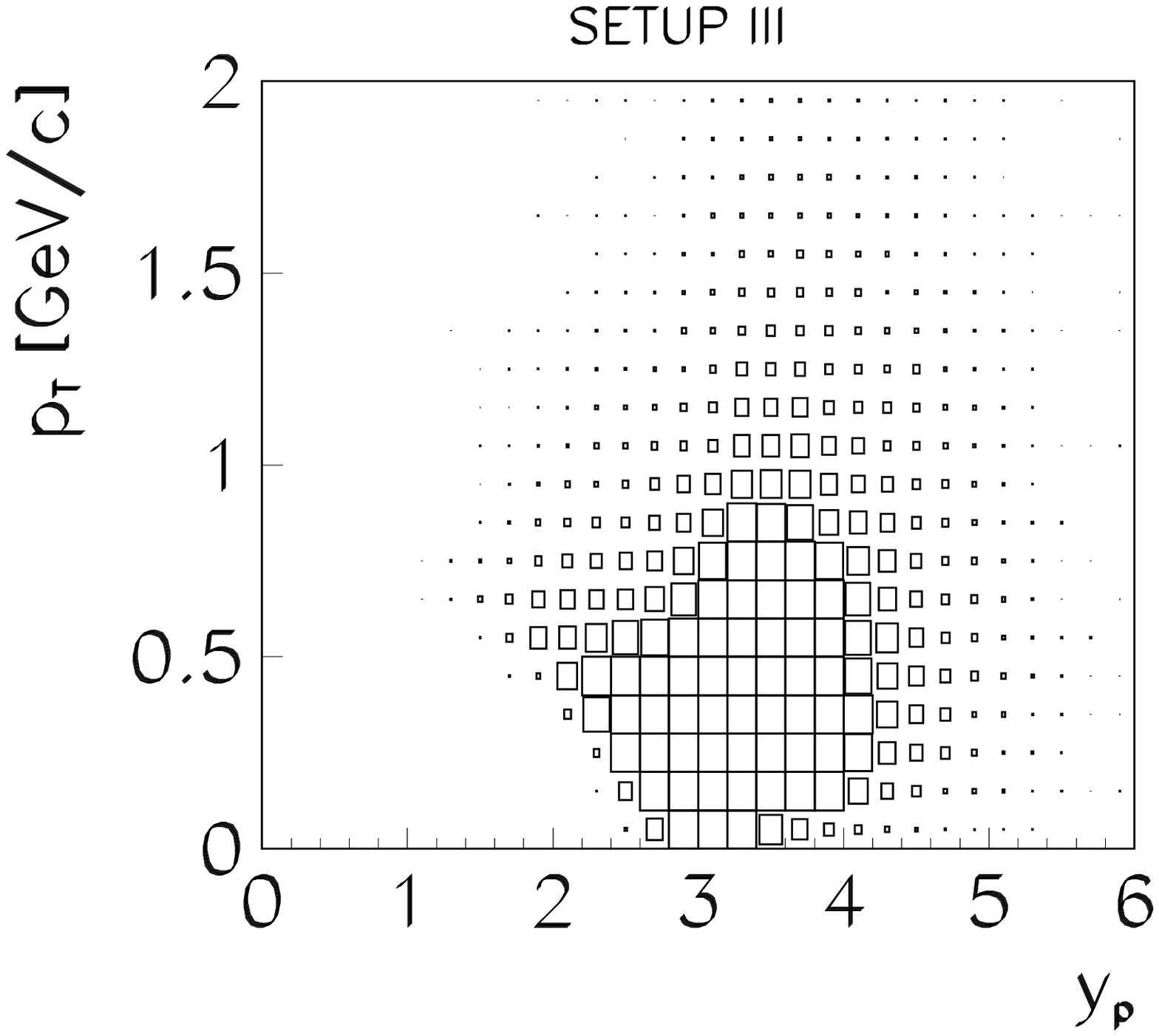,height=8.0cm,width=8cm}}
\end{minipage}
\vspace{-1.5cm}
\caption{Acceptance of the streamer chamber in configuration SETUP
II (left) and SETUP III (right) for protons,
illustrated by the phase space population in a typical heavy ion
collision. Note the forward acceptance of SETUP III.}
\label{acc_p}
\end{figure}

In order to measure forward-going strange particles and 
participant protons in
the projectile hemisphere for the heaviest system
(S+Au), the target and a sweeper magnet (SM) were installed upstream
of the standard SETUP I configuration as shown in 
Fig.~\ref{sweeper} (called SETUP III). The sweeper magnet has a vertical
entrance opening of $\pm$2.5$^{\circ}$ for particles emerging from
the target placed 6.3 cm upstream of this magnet. Due to its bending 
power of about 4.5 Tm 
all low momentum charged particles are swept out of the
acceptance of the second magnet thus
reducing the track density in the streamer chamber. 
This configuration not only enables measurement of strange particle
decays in the projectile hemisphere but also covers 
participant protons beyond midrapidity ($y_{cm}=3.0$) up to beam rapidity 
($y_{proj}=6.0$). This
can  be seen in Fig.~\ref{acc_p} where the population of protons in
transverse momentum $p_T$ and rapidity $y$ is presented. 
When sulphur projectiles are used the proton acceptance can be extended up to beam rapidity
only in SETUP III.

\begin{figure}[H]
\centerline{\epsfig{file=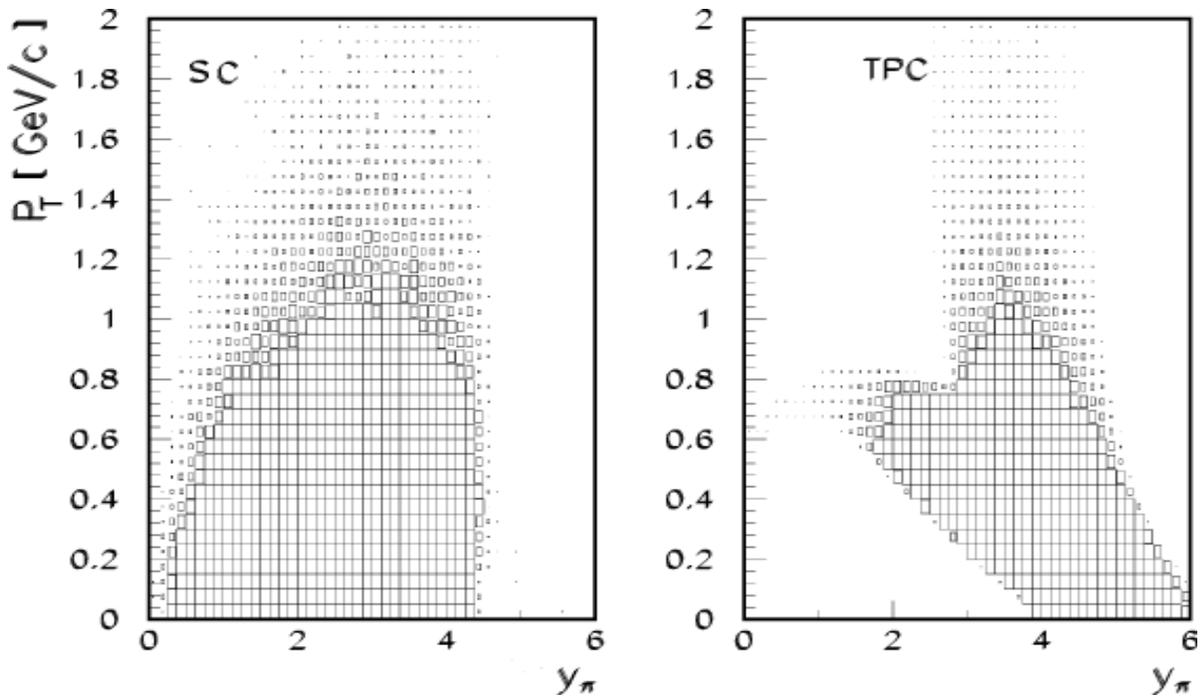,height=7.0cm,width=15.5cm}}
\vspace{0.1cm}
\caption{Acceptance of the two large-volume tracking detectors,
the streamer chamber (left)
and TPC (right) for SETUP II, 
illustrated by the phase-space population of negatively
charged hadrons in a typical heavy ion collision.}
\label{acc}
\end{figure}

A second large-volume tracking device, a time
projection chamber with a sensitive volume  of 240 $\times$ 
125 $\times$ 112 cm$^3$ \cite{TPCNIM} located downstream 
behind the analysing magnet was installed in 1990 and utilized 
in the $^{32}$S--beam running for all three setups. It was also used 
in the deuteron--nucleus studies in the SETUP I configuration.
The TPC detects charged particles by measuring the ionisation
of a particle, traversing the TPC, up to 60 times. 
For a primary track emerging from the production vertex the
momentum of the particle can be calculated from the direction and
position of the track 
inside the TPC. The TPC was positioned for most of
the time such that it detected magnetically stiff negatively charged
particles above midrapidity. 
The acceptance for negatively charged hadrons (pions) 
is illustrated in Fig.~\ref{acc}. Both tracking detectors combined 
cover almost all momentum space available for hadron production
at SPS energy.
In addition, the TPC identified charged particles by multiple sampling
of the specific ionisation. This capability was used to measure
antiproton spectra \cite{Pbar} and to distinguish electrons from
negatively charged hadrons.

%
%
\newpage

\newpage

\section{Data sets}

NA35 has studied a variety of collision systems, ranging from 
minimum bias proton--sulphur interactions to central sulphur--gold
reactions. 
The central-collision fractions of the inelastic cross sections of
the nucleus--nucleus 
collisions are typically the upper
few percent of the inelastic cross section for the 
heavy ion induced reactions (see Table~\ref{systems}).
Central events are selected either by a minimal energy deposition
in the projectile fragmentation domain at forward angles
($\Theta_{lab} \le 0.3^{\circ}$) which is covered by the veto calorimeter
(FET trigger), or by a
large transverse energy deposition in the midrapidity calorimeters
(TET trigger).
In the case of p+S interactions a trigger  was derived
from a scintillation counter placed downstream of the streamer chamber which
covered the projectile fragmentation domain. Its veto threshold was set
just below the signal of the projectile particle  
and it selected essentially all inelastic interactions. 
An additional off-line selection of events with a charged particle 
multiplicity larger than five was applied in order
to exclude contaminations due to non-target interactions in 
low multiplicity events. 


\begin{table}[H]
\caption{Characteristics of the data sets: 
colliding system, experimental configuration, number of events, trigger type,
the selected fraction of the inelastic cross section, 
acceptance in rapidity for negatively charged hadrons and for 
protons. (The trigger for the d+Au interactions marked by ($^*$) 
selects collisions where
the deuteron projectile fully overlaps with the gold target.)} 
\label{systems}
\begin{center}
\begin{tabular}{|c|c|c|c|c|c|c|}
\hline   &   &   &   &   &   &   \cr 
 Reaction  &
 Setup   &
 Number & 
 Trigger &
 $ \frac {\sigma} {\sigma_{inel.}} $  [$\%$]  &
 Acceptance &
 Acceptance \cr
         &   & of events  &   &   & $h^-$  & protons  \cr 
\hline
\hline
  p+S  &  SETUP I          &     & scint. veto &    &      &      \cr
       &                   &     & $h^{+-}>5$  &      &  & \cr
       &  internal target  & 735 &             & 63.5 & $-2 \leq y_{\pi} \leq
4$ & $-1 \leq y_{p} \leq 3$ \cr
       &  external target  & 339 &             & 63.5 & $0.6 \leq y_{\pi}
\leq 7$ & $0.2 \leq y_{p} \leq 6$ \cr
\hline
  p+Au &  SETUP I          & 1603&    FET    & 67 & $0.6 \leq y_{\pi} 
\leq 6$ & $0.2 \leq y_{p} \leq 6$ \cr
\hline
  d+Au &  SETUP I          & 2457&   FET$^*$ & 43 & $0.4 \leq y_{\pi}
\leq 6$ & $0.2 \leq y_{p} \leq 6$ \cr
       &  TPC              & 9484&   FET$^*$ & 43 &   -    & $2.8 \leq y_{p}
\leq 4.4$ \cr
\hline
  O+Au &  SETUP I          & 565 &    FET    & 1.8 & $0.6 \leq y_{\pi}
\leq 3.8$ & $0.2 \leq y_{p} \leq 3$ \cr
\hline
  S+S  &  SETUP II         & 2650&    FET    & 2.9 & $0.6 \leq y_{\pi}
\leq 3.8$ & $0.2 \leq y_{p} \leq 3$ \cr
       &  TPC              &23500&    FET    & 3.5 & $3.2 \leq y_{\pi}
\leq 5.4$ & - \cr
\hline
  S+Ag &  SETUP II         &13730&    FET    & 3.2 & $0.6 \leq y_{\pi}
\leq 3.8$ & $0.2 \leq y_{p} \leq 3$ \cr
       &  TPC              &18000&    FET    & 3.2 & $3.2 \leq y_{\pi}
\leq 5.4$ & - \cr
\hline
  S+Au &  SETUP I         & 331 &    TET    & 1.3 & $0.8 \leq y_{\pi}
\leq 3.6$ & - \cr
       &  SETUP III        & 3142&    FET    & 6.3 & $3.8 \leq y_{\pi}  
\leq 6.0$ & $2.8 \leq y_{p} \leq 6$ \cr
       &  TPC              &25700&    FET    & 6.3 & $3.2 \leq y_{\pi}
\leq 5.4$ & - \cr
\hline
\end{tabular}
\end{center}
\end{table}

\newpage

\section{Data Analysis}

The operation of the streamer chamber (SC) is described elsewhere 
\cite{Bri95,SCNIM}, 
the data analysis will briefly be described in the following.
Streamer chamber pictures were scanned and measured using two different 
approaches: using the traditional film measurement by human operators
or applying image processing and pattern recognition techniques to
digitized streamer chamber data. 
Most of the streamer chamber data were analyzed automatically 
by the Frankfurt Image Processing System which is described in detail
elsewhere \cite{Bri95}. The
three perspective views of a streamer chamber event were
recorded on 70 mm film and later digitized by a high resolution
CCD-linescan camera. These digitized images were submitted to a
series of image processing and pattern recognition programs.
Tracks matched in the three stereoscopic views were then
reconstructed in space. The particle momenta and the production
vertex were obtained by fitting a trajectory to the measured points.
The low multiplicity data obtained in SETUP I were analyzed on
traditional projection tables by human operators.
 
Since the time projection chamber (TPC) was new to NA35 and
relatively new  for heavy ion applications \cite{TPCNIM}, 
we will describe its operation briefly. 
A charged particle which traverses the sensitive volume of the TPC,
ionizes gas molecules and produces electrons along its trajectory.
Guided by a homogeneous electric drift field,
the electrons drift towards the
read-out MWPCs of the TPC. A field gradient at the
sense wires
leads to a multiplication of the drifted electrons by
$5 \cdot 10^3$. The resulting space charge induces a
signal onto
the segmented cathode plane (pads). 
The centroid of these induced signals (clusters) can be determined 
with an accuracy
of $\sigma = 500$ $\mu$m in pad direction and of $\sigma = 310$ $\mu$m in 
drift (time) direction. Pattern recognition is done by a simple
track follower, which associates clusters with a straight line
\cite{Roland}. 
The particle momenta are determined by assuming that the particles emerge
from a known production vertex. 

\subsection{Track recognition and reconstruction efficiency}

The momentum resolution achieved with the streamer chamber as
a continuous tracking device is approximately 3 MeV/c 
($dp/p = 0.15\%$) for a typical pion 
with momentum of 2 GeV/c. The  momenta in the TPC 
are typically 
higher than in the streamer chamber, since  the TPC is located at more 
forward angles in  the laboratory. The TPC resolution for 10 GeV/c tracks
is approximately $ 1\%$.


In the streamer chamber track reconstruction losses due
to tracks which overlap in the individual camera projections as well as 
to two-track resolution, pattern
recognition and space reconstruction inefficiencies can be
examined by either of two methods. 
The first method relies on  human  
visual scanning of film data and multiple independent
scans of the same data.
The second method is a Monte Carlo simulation of digitally processed images.
By this method heavy ion events are simulated using the FRITIOF event 
generator code \cite{frit}. The track trajectories
are modeled through the NA35 experimental set-up using the GEANT
detector simulation package \cite{gean}. For each track the streamers 
are calculated and then projected through the 
imaging system onto film. 
The film response and digitisation process are then simulated.
The resulting digital event pictures are
submitted to the entire pattern recognition and track reconstruction
process.
This procedure provides the efficiency as a function of
the
two kinematical variables rapidity $y$ and transverse momentum
$p_T$. 
Both methods give an
overall efficiency of about 95-97$\%$ for the highest multiplicities
in our acceptance.

For the TPC the reconstruction efficiency was determined by a
Monte-Carlo simulation using the VENUS event generator
\cite{Venus}, the GEANT-program and a simulation of the
TPC response \cite{Appel}. The efficiency was found to be
90-95$\%$, approximately independent of $y$ and $p_T$ within the
geometrical acceptance. For the particle spectra the inefficiencies were
corrected as a function of $y$ and $p_T$.

\begin{figure}[hbt]
\centerline{\epsfig{file=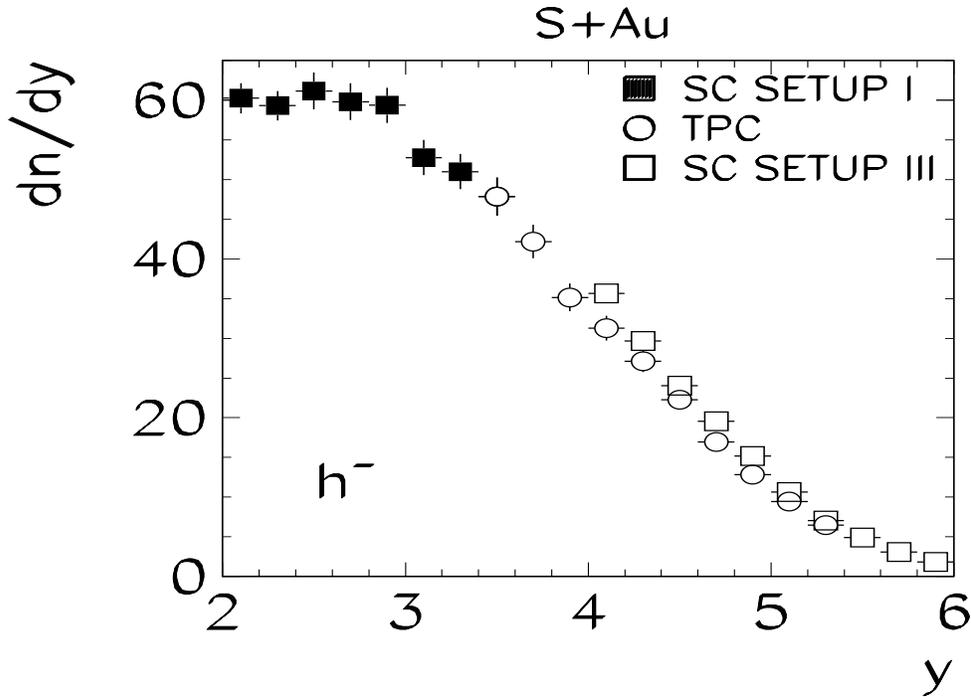,height=10cm,width=14cm}}
\caption{Rapidity distributions of negatively charged hadrons
(pion mass assumed) produced in central S+Au collisions, 
measured in the streamer chamber
(configurations SETUP I and SETUP III) and in the TPC.}
\label{tpc_sweeper}
\end{figure}

\begin{figure}[hbt]
\centerline{\epsfig{file=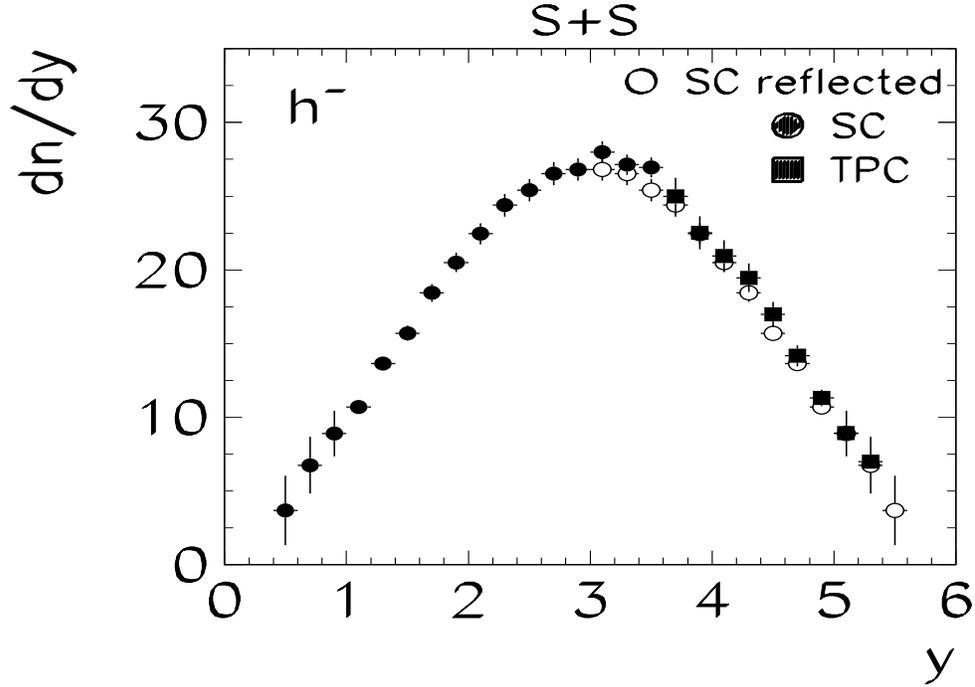,height=10cm,width=14cm}}
\caption{Rapidity distributions of negatively charged hadrons
produced in central S+S collisions, measured in the streamer chamber
(SETUP II) and reflected around $y_{cm}=3$,  and in the TPC.}
\label{tpc_sweeper_2}
\end{figure}

A comparison of rapidity densities of negatively charged hadrons obtained
by the streamer chamber analysis of SETUP II and SETUP III data  and
by the TPC analysis is shown in Fig.~\ref{tpc_sweeper}. For the
symmetric system S+S streamer chamber data below midrapidity can
be reflected about $y_{cm}=3.0$ and then compared to TPC data
above midrapidity as shown in Fig.~\ref{tpc_sweeper_2}.   
The TPC data below $y = 4.0$ and above $y = 4.6$ have been
extrapolated to low and high $p_T$ respectively by an exponential in 
$1/p_T \cdot dn/dp_T$.   
The comparisons in both figures demonstrate 
a satisfactory agreement (within several $\%$).

%
%
%
%

\newpage
\subsection{Particle identification}

We define negatively charged hadrons $h^-$ as the sum of $\pi^-$, $K^-$ and
$\overline{p}$. Their average multiplicities in central S+S collisions
are estimated to be about 90:7:1 \cite{Alb94,Pbar}. Corrections for electron 
contamination were made on a statistical basis (see next chapter).  
The rapidity of negatively charged hadrons was calculated assuming
the pion mass, throughout; this implies that the $K^-$ and $\overline{p}$ contributions
to $h^-$ rapidity distributions appear with an upward shift relative to their 
true position. A slight asymmetric widening of the rapidity distribution results.
Since the streamer chamber has very little particle identification 
capability for stable
particles, the net proton distribution was deduced from a
measurement of the charge excess
\[ \frac{dn}{dy}(AA' \rightarrow (p-\overline{p})) \approx
        \frac{dn}{dy}(AA' \rightarrow h^+) - \frac{dn}{dy}(AA'
\rightarrow h^-),  
\]
where the rapidity is calculated assuming the proton mass.
This is a good approximation for all rapidities 
in collisions of isoscalar nuclei (equal number of neutrons and protons) 
assuming local isospin conservation. Even 
for asymmetric systems deviations are expected to be 
small. Equating net protons to the charge-excess is
subject to corrections which will be discussed below.

\subsection{Corrections}

All data in this paper are corrected for geometric acceptance and
reconstruction inefficiencies, for $e^+$ and $e^-$
misidentified as hadrons, for hadrons resulting from the weak
decays of $K^0_S$, hyperons
and for hadrons produced in 
secondary hadron-nucleus interactions.

\begin{figure}[hbt]
\centerline{\epsfig{figure=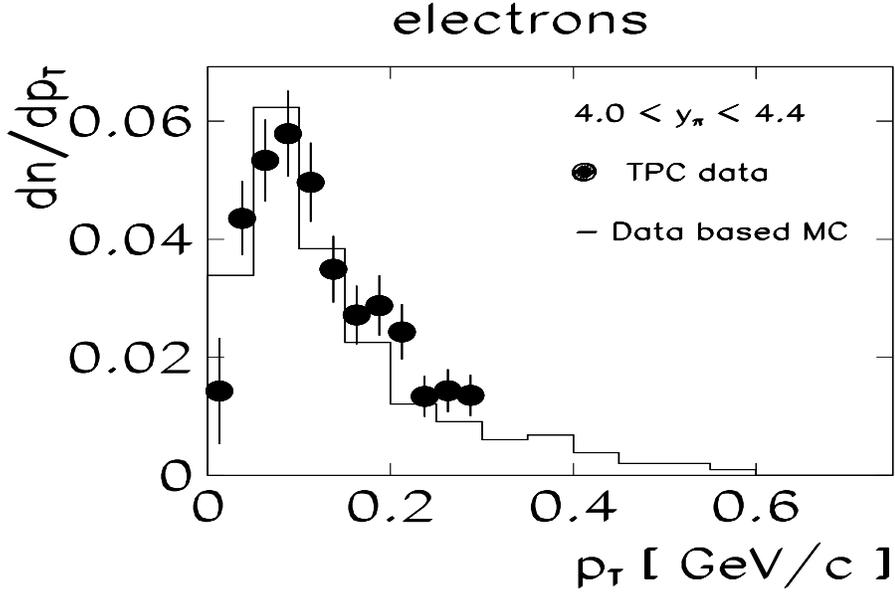,height=8cm,width=12cm}}
\caption{Transverse momentum spectrum of conversion electrons in central
S+S collisions
measured and identified in the TPC in the 
pion rapidity range $4.0 \leq y \leq 4.4$ compared to a calculated
spectrum based on the measurement of negatively charged hadrons
(see text for details). The vertical scale is given in (GeV/c)$^{-1}$.}
\label{electron}
\end{figure}

One source of contamination in the charged hadron spectra is the 
2$\gamma$-decay of 
the abundantly produced $\pi^0$ inside the target and subsequent conversion 
of $\gamma$s into $e^+e^-$--pairs. 
The spectral 
distributions of contaminant electrons were obtained from a Monte Carlo 
calculation using
the experimental negative particle distributions and assuming equal
$\pi^0$ and $\pi^-$ meson multiplicities and spectra.
An alternative method is to use the $\pi^0$ spectra obtained from
an event generator (FRITIOF), with parameters tuned to reproduce
the measured spectra, and a GEANT simulation for the subsequent decays and
conversions. The latter method was employed
at high rapidities, where the data based Monte Carlo may suffer from
reconstruction inefficiencies at high momenta. The methods give
similar results, which are confirmed by identified electron spectra that are 
measured in the TPC and displayed in Fig.~\ref{electron}. 
The electron contribution to the charged particle spectra is
subtracted for each $(y,p_T)$ bin.  
 
The contamination of the negatively charged hadron spectra by
pions produced in secondary interactions primarily in the target, and
in weak decays 
is corrected for
on the basis of GEANT simulations (see e.g. \cite{Teitel}). The parameters
of the event generators were tuned to reproduce the measured
strange particle yields and spectra.  

\begin{figure}[H]
\centerline{\epsfig{figure=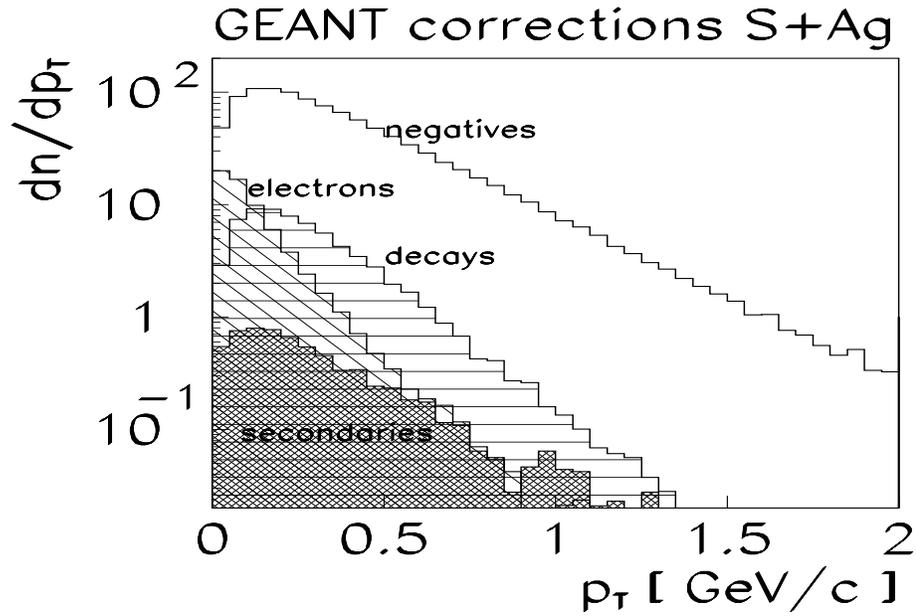,height=8cm,width=12cm}}
\vspace{-1cm}
\caption{Composition of a measured transverse momentum spectrum of
negatively charged particles, based on a tuned event generator
(FRITIOF) and a GEANT simulation. The vertical scale is given in (GeV/c)$^{-1}$.}
\label{geant_pt}
\end{figure}

\begin{figure}[H]
\centerline{\epsfig{figure=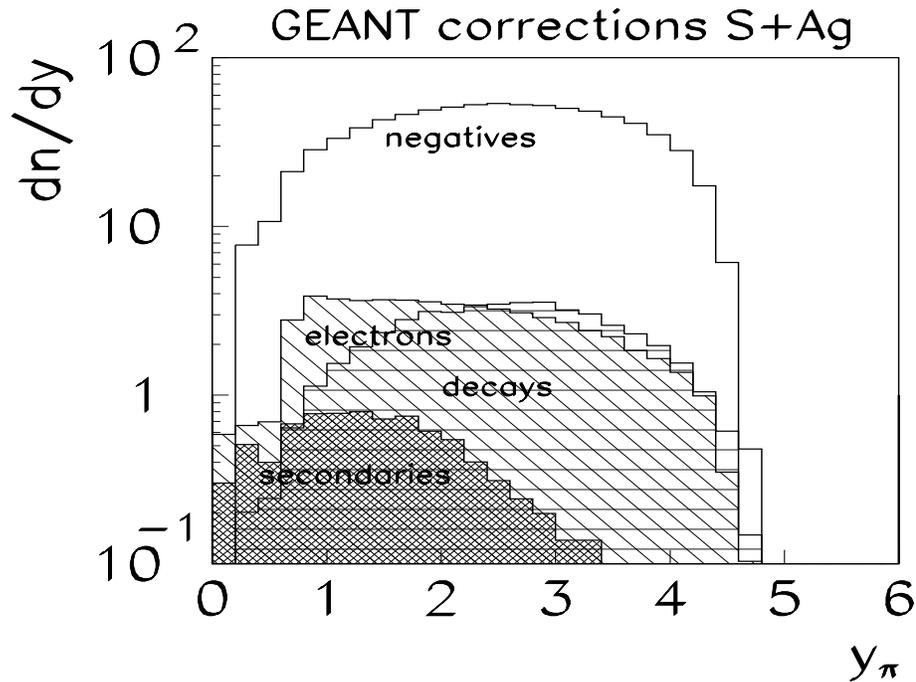,height=9cm,width=12cm}}
\vspace{-1cm}
\caption{Composition of a measured rapidity spectrum of negatively
charged particles, based on a tuned event generator
(FRITIOF) and a GEANT simulation.}
\label{geant_y}
\end{figure}

The contributions of the various contaminations 
as functions of $p_T$ and $y$ for a central S+Ag collision are shown in
Figs.~\ref{geant_pt} and \ref{geant_y} respectively.
The contaminations are largest at small $p_T$. The electrons 
dominate below transverse momenta of 150 MeV/c. 
The first $p_T$ bin (50 MeV/c) especially is strongly affected by the
electron correction, the amount of the contamination can be up to 50\%. 
Above 150 MeV/c the weak decay products represent the main 
contamination to the spectrum. 
For larger transverse momenta the various contributions get smaller. 
The rapidity dependence of the decay product contamination 
follows the distribution
of the primaries, while electrons and secondary interactions contribute
more strongly at low rapidities (Fig.~\ref{geant_y}).    
The overall systematic error on the spectra is estimated to be less than
5\% for $p_T$ above 50 MeV/c. 

All data - spectra of both negatively and positively charged hadrons - 
were corrected as described above. 
In addition, when deriving the net proton spectra from the 
charge excess, a
correction is applied for the excess of the
$K^+$ relative to the $K^-$ yield. 
Since these corrections
are based on measurements of the same experiment over a wide
acceptance, the corrections are model independent.
Nevertheless, these corrections become large at low transverse 
momenta, particularly at low rapidities, and therefore a low-$p_T$ cut
was applied to the net proton spectra. This cut also takes care of possible 
differences in the $\pi^+$ and $\pi^-$ yields for non-isoscalar systems
of colliding nuclei. In order to obtain rapidity densities
the transverse momentum spectrum 
was extrapolated into the part below the cut
using a thermal fit to the data. This extrapolation contributes 
approximately 20\% (assuming a temperature of about 200 MeV and a 
$p_T$-cut of 300 MeV/c) to the cross section. 

\begin{figure}[H]
\centerline{\epsfig{file=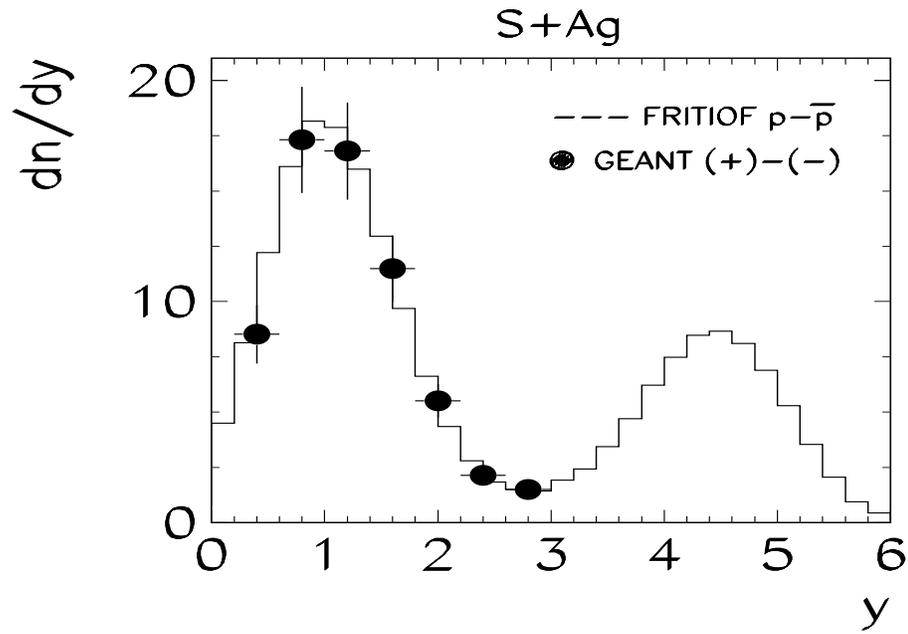,height=9cm,width=13cm}}
\vspace{-1cm}
\caption{Simulated rapidity distributions of net protons 
($p-\overline{p}$) and of reconstructed 
net protons deduced from the charge excess after all corrections 
for central S+Ag collisions.}
\label{58}
\end{figure}

The validity of equating participant protons to the charge-excess even
in asymmetric systems is demonstrated in Fig.~\ref{58}. 
Particles taken from a simulation
(FRITIOF) were passed to a GEANT-program of the experimental setup. 
After all corrections described above the net proton distribution 
which was deduced from the charge excess agrees very well with 
the input distribution of net protons within the acceptance $0.2 < y < 3.0$. 
The overall systematic error on the participant proton
spectra is estimated to be less than 12\%. Note that this conclusion
does not rest on the implication that the FRITIOF prediction in
Fig.~\ref{58} gives a good account of the data (see sect. 6.1.3).


\newpage
\newpage
%
%

\newpage

\section{Net Baryons}


In a simple picture of high energy nucleus--nucleus collisions
one distinguishes two classes of nucleons: spectators and participants.
Spectator nucleons are defined as those  which did not suffer 
collisions and therefore have their momenta located inside the
Fermi spheres of the interacting nuclei.
Participant nucleons
are those which suffer collisions that affect their momentum 
sufficiently to remove them from the Fermi sphere of the 
incoming nuclei. Some of them may be converted into strange and
multistrange baryons.
Integrating the rapidity distributions of net baryons (baryons minus antibaryons)
between $y = 0.2$ and $y = 5.8$ yields an estimate of the total number 
of participant baryons. 
This estimate is only accurate in the case of central collisions of
identical nuclei (S+S), 
where the number of spectators is small compared to the  number of
participants so that the result of the technical procedure to
separate spectators and participants depends only weakly on 
the cuts used. For asymmetric systems the net baryon number 
as calculated in this paper should be treated as a crude approximation of the 
number of participants since cascading processes in the target spectator 
matter may invalidate the definition of a spectator.


\subsection{Rapidity distributions}

Rapidity distributions of net protons were obtained from the
charge excess with the associated corrections discussed above. Since
these corrections become large at low transverse momenta, a cut
was applied at 300 MeV/c (for S+A collisions at 400 MeV/c). 
The missing low momentum part 
was extrapolated on the basis of a thermal fit in order 
to obtain rapidity densities (see chapter 5.2). In comparing
rapidity distributions for the various systems one has to keep in mind
that different forward energy triggers may affect the spectra differently.

\subsubsection{Proton-Nucleus Collisions}

\begin{figure}[H]
\vspace{-1cm}
\centerline{\epsfig{file=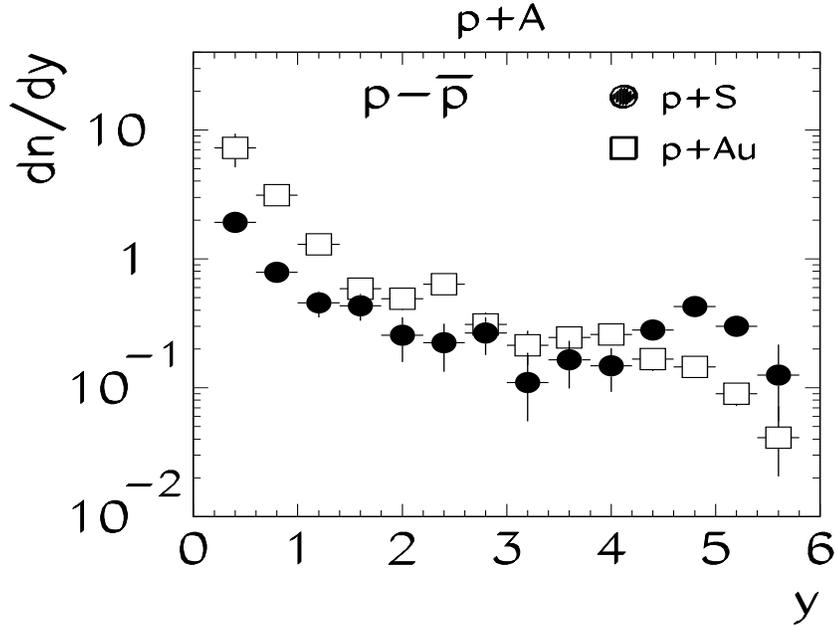,height=9cm,width=12cm}}
\vspace{-1cm}
\caption{Rapidity distributions of net protons ($p-\overline{p}$)
for minimum bias p+S and p+Au collisions at 200 GeV/nucleon.}
\label{60}
\end{figure}

The rapidity distributions of net protons for minimum bias
p+S and p+Au interactions are shown in Fig.~\ref{60}. While the
rapidity densities around midrapidity ($y_{cm} = 3)$ 
for both reactions are not very different,
clear differences are observed for rapidities below 1.2 and larger than 4.4.
In p+Au collisions more target nucleons participate in the reactions
and are therefore shifted up to one unit in rapidity farther than in p+S 
collisions. 
For the projectile
the gold nucleus looks black, i.e. the probability of the projectile
to traverse the target nucleus without losing any or little energy is
small. 
For p+S interactions on the other hand, one can perceive a clear 
projectile proton peak at rapidity $\approx$ 4.7.

%


\subsubsection{Nucleus-Gold Collisions}

\begin{figure}[H]
\centerline{\epsfig{file=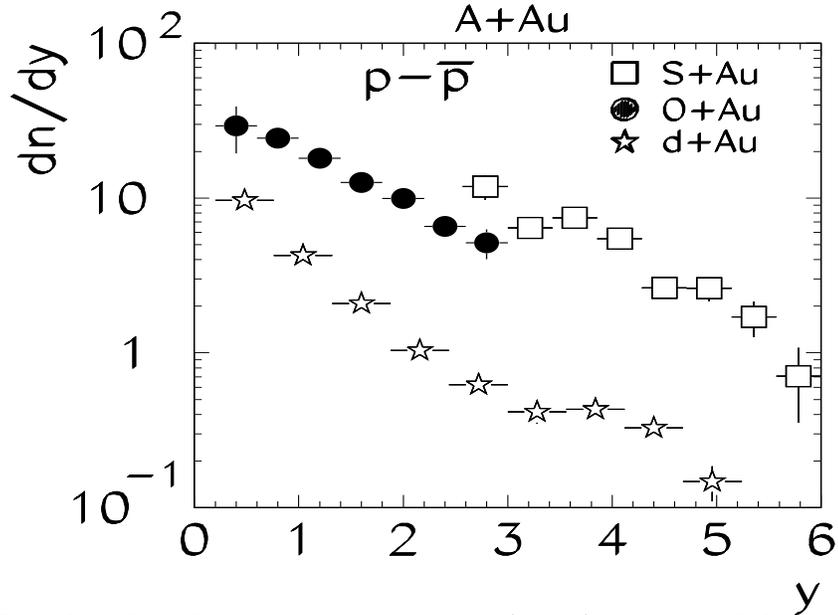,height=9cm,width=12cm}}
\vspace{-1.5cm}
\caption{Rapidity distributions of net protons ($p-\overline{p}$)
for central d+Au, central O+Au and central S+Au collisions at 200 GeV/nucleon.}
\label{61}
\end{figure}
\begin{figure}[H]
\centerline{\epsfig{file=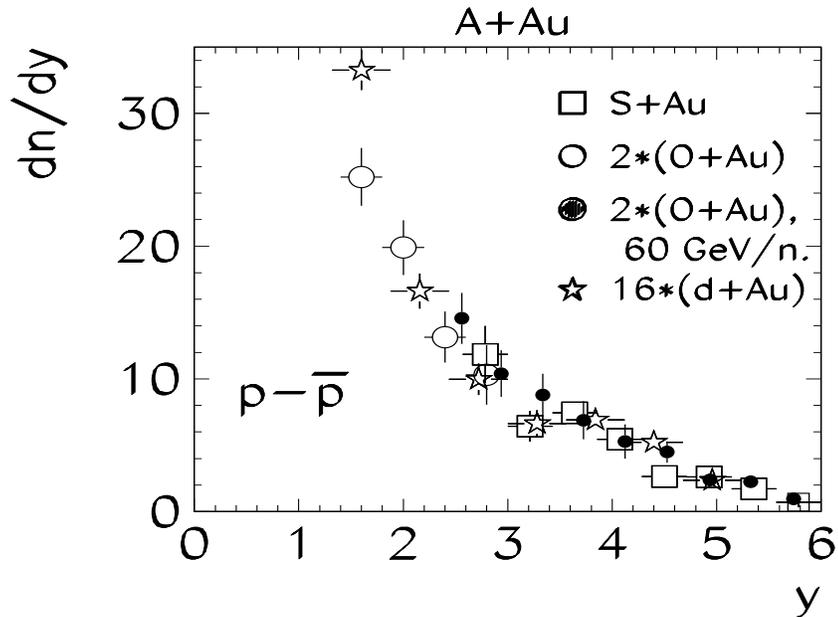,height=9cm,width=12cm}}
\vspace{-1.5cm}
\caption{Scaled rapidity distributions of net protons 
for central $16 \cdot$(d+Au), central $2 \cdot$(O+Au) and 
central S+Au collisions. 
For the case of the 
60 GeV/nucleon data the rapidities are scaled by the ratio of the beam
rapidity relative to that of 200 GeV/nucleon.}
\label{70}
\end{figure}

In order to study the dependence of nuclear stopping on the mass of the
projectile in the case of incidence on a heavy target nucleus, 
net proton rapidity
distributions were measured for various `full overlap' collisions: 
central d+Au, central O+Au and central
S+Au reactions. These data are displayed in Fig.~\ref{61}. 
O+Au collisions were
measured in the streamer chamber for rapidities below midrapidity, 
$y < y_{cm}$, SETUP I. 
The S+Au data were measured in the streamer chamber configuration SETUP III 
at rapidities
$y > y_{cm}$. The rapidity density at midrapidity, $y = y_{cm}$, is observed to
increase with projectile mass. A comparison of 
the rapidity distribution for S+Au collisions with the distributions for 
d+Au and O+Au, multiplied by 16 and 2 
(the inverse ratios of the projectile nucleons relative to
the S+Au reaction), respectively, is displayed in Fig.~\ref{70}. 
For the O+Au data 
at a beam energy of 60 GeV/nucleon
the rapidity axis is scaled by the
ratio of the beam rapidities 
($y_{proj}$ (200 GeV/nucleon) / $y_{proj}$ (60 GeV/nucleon) = 6.0/4.8) 
prior to plotting in Fig.~\ref{70} \cite{Tonse,Roe1}. 
The resultant rapidity distributions are similar 
for $y > y_{cm}$. 
This implies that the stopping of projectile nucleons
is rather independent of the size of the projectile nucleus and its energy 
in collisions with a heavy target nucleus. 

%


\subsubsection{Sulphur-Nucleus Collisions}

\begin{figure}[H]
\centerline{\epsfig{file=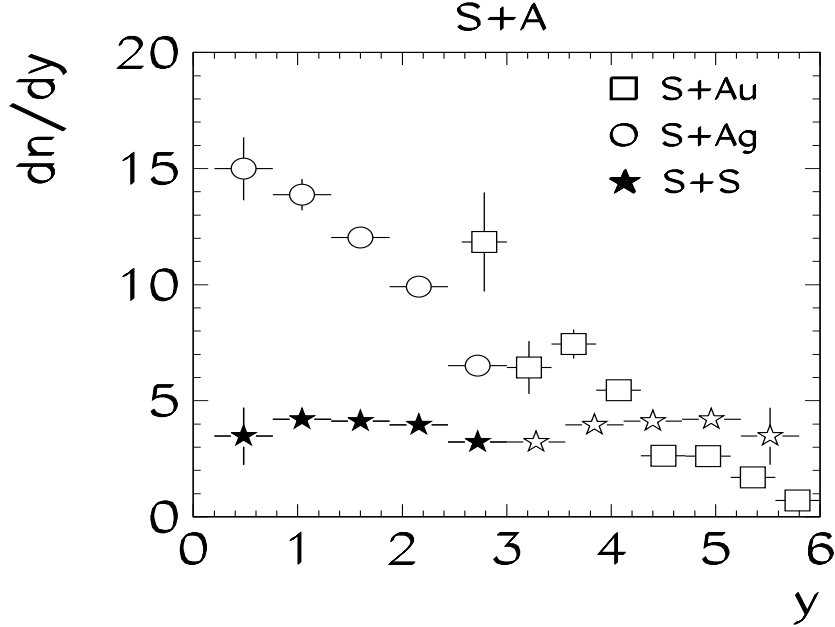,height=9cm,width=12cm}}
\vspace{-1.0cm}
\caption{Rapidity distributions of net protons ($p-\overline{p}$)
for various central S--nucleus collisions at 200 GeV/nucleon.}
\label{62}
\end{figure}

To investigate the effect of changing the nuclear thickness of the target,
rapidity distributions of net protons were measured for 
central S+S, S+Ag and S+Au collisions. The data are shown in Fig.~\ref{62}.
The S+S and S+Ag data were measured in the SETUP II
for $y < y_{cm}$ and 
S+Au data in the SETUP III for $y > y_{cm}$. 
For the symmetric system S+S, 
the data points reflected at $y_{cm}$ are also shown. 
The shape of the net proton rapidity distribution changes from 
being relatively flat for the
light symmetric system (S+S) to a monotonic decrease with
rapidity for the 
asymmetric systems.
The rapidity density at $y = y_{cm}$ increases for heavier projectiles,
and is a factor of about two larger for S+Au relative to S+S collisions.
There is an excess of net protons in the forward direction for the lighter
S+S reaction relative to the S+Au reaction. Thus, it is evident that the 
S target
nucleus is less effective than the Au nucleus in stopping (slowing down)
incident projectile nucleons. For the S+Au system, projectile protons are
shifted further backward in rapidity than for the S+S system.
The rapidity densities near midrapidity agree with the values obtained by
the NA44 experiment for identified protons and antiprotons as shown in 
Table~\ref{NA44}. 
\begin{table}[hbt]
\caption{Rapidity densities of net protons 
($p - \overline{p}$) near midrapidity ($y \approx 2.8$) in central S+S and 
S+Au(Pb) collisions \protect\cite{Dodd}. } 
\label{NA44}
\begin{center}
\begin{tabular}{|c|c|c|}
\hline   
 & \multicolumn{2}{|c|}{ $dn/dy$  $(p-\overline{p})$} \cr
\hline
 Reaction &
 this paper &
 NA44 \protect\cite{Dodd}  \cr 
\hline
\hline
  S+S  & 3.2 $\pm$ 0.4 &  3.2 $\pm$ 1.5   \cr
\hline
  S+Au(Pb) &  12 $\pm$ 2  &  12.4 $\pm$ 2.1 \cr
\hline
\end{tabular}
\end{center}
\end{table}


\subsubsection{Rapidity Shift}

\begin{figure}[H]
\centerline{\epsfig{file=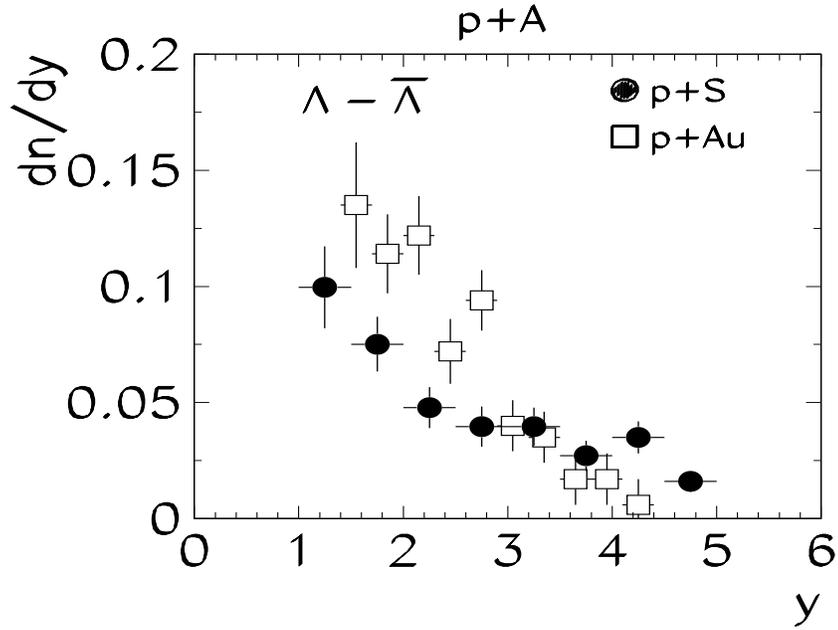,height=9cm,width=12cm}}
\vspace{-1.0cm}
\caption{Rapidity distributions of net hyperons
(${\Lambda-\overline{\Lambda}}$) for
minimum bias p+S ($1.0 < y < 5.0$) and p+Au ($1.4 < y < 4.4$) 
collisions at 200 GeV/nucleon.}
\label{72}
\end{figure}

\begin{figure}[H]
\vspace{-2.0cm}
\centerline{\epsfig{file=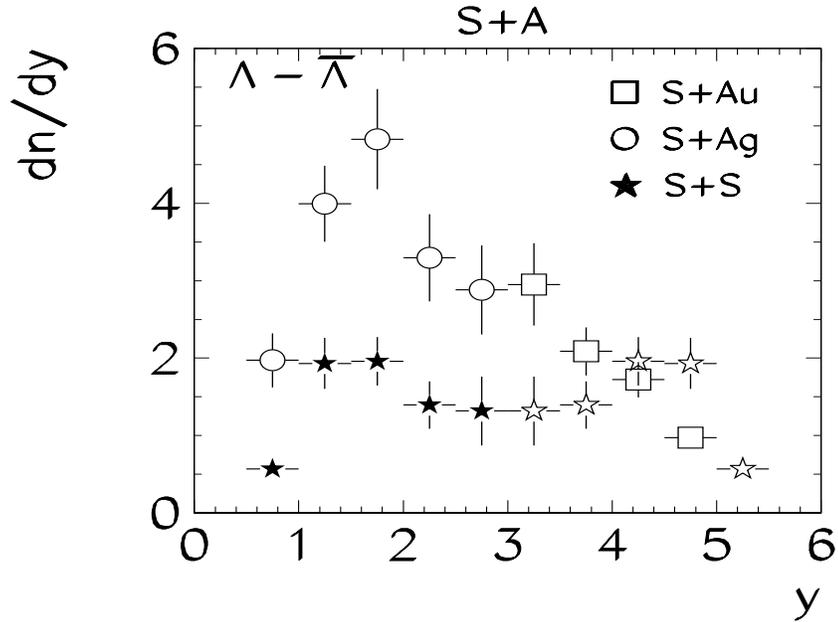,height=9cm,width=12cm}}
\vspace{-1.0cm}
\caption{Rapidity distributions of net hyperons
(${\Lambda-\overline{\Lambda}}$) for
central S+S ($0.5 < y < 3.0$), S+Ag ($0.5 < y < 3.0$) and 
S+Au ($3.0 < y < 5.0$) collisions at 200 GeV/nucleon. The solid stars
are measured S+S data, the open stars are reflected at $y = y_{cm}$.}
\label{71}
\end{figure}

Before the collision the nucleon rapidity distribution is
concentrated in peaks at the target rapidity $y = 0$ and at 
beam rapidity $y = y_{proj}$,
with the peak width given by the Fermi momentum inside the nuclei
($\Delta y \approx 0.2$). 
After the collision, 
target and projectile nucleons are shifted towards midrapidity, with some
nucleons converted into (multi)strange baryons. The latter must also 
be counted in the evaluation of rapidity loss.  
The rapidity distributions of net hyperons 
($\Lambda-\overline{\Lambda}$) for minimum bias p+S and
p+Au collisions are displayed in Fig.~\ref{72}. 
Those for central collisions of S+S, S+Ag, and S+Au are displayed in 
Fig.~\ref{71}. The trends in the rapidity
distributions for net hyperons as a function of target nucleus are similar to
those for net protons described in the previous section.

In order to quantify the amount of stopping in symmetric and
asymmetric collisions, the average rapidity loss  
(or mean rapidity shift) $\langle \delta y \rangle 
= y_{proj} - \langle y \rangle$ is defined, 
where $y_{proj}$ is the incoming projectile rapidity and $\langle y \rangle$
is the average net baryon rapidity after the collision.
For symmetric systems the average rapidity was calculated by
averaging the rapidity of the net protons in the interval from $y_{cm} = 3$ to 
$y_{proj} - 0.2$.
For the asymmetric systems $\langle y \rangle$ is
defined as the average rapidity of the net proton $dn/dy$ distribution
in the projectile hemisphere from $y_{proj} - 0.2$ to a rapidity
such that the integral under the distribution equals the number of
net projectile protons \cite{Hansen}. 
If the target and projectile nucleons mix, 
the rapidity loss defined in this manner is a lower limit. 

\begin{figure}[H]
\centerline{\epsfig{file=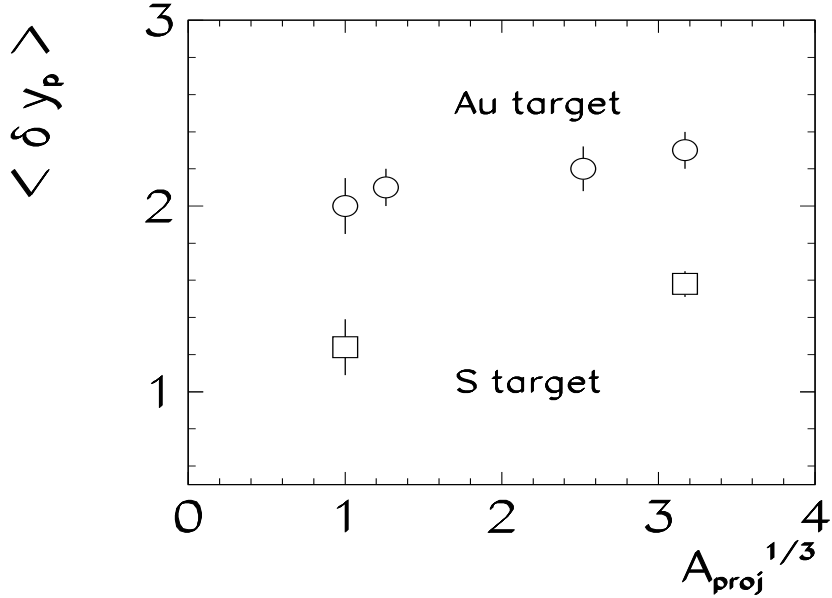,height=9cm,width=12cm}}
\caption{Average rapidity loss (mean rapidity shift) of net 
projectile protons 
for various projectiles incident on sulphur and gold targets at 
200 GeV per nucleon.}
\label{rapshift}
\end{figure}

The average rapidity losses $\langle \delta y \rangle$ for collisions of various
projectiles with sulphur and gold targets are displayed in Fig.~\ref{rapshift}. 
For each
target $\langle \delta y \rangle$ is approximately independent of projectile mass (but
may also be construed as being consistent with a slight increase with
projectile mass). The most obvious trend is a shift from 
$\langle \delta y \rangle$ = 1.2 - 1.6 for
sulphur target interactions to $\langle \delta y \rangle$ = 2.0 - 2.3 
for the gold target interactions.


\subsubsection{Number of Net Baryons}

By integrating the rapidity distributions the number of net
protons inside the acceptance was obtained and is shown in 
Table~\ref{participant}. 
The total number of net protons
is defined as the
net proton number in the final state
outside the `Fermi spheres'
of the projectile and target nuclei - approximated by selecting a 
rapidity range 0.2 to 5.8.
For systems where the
acceptance does not cover the rapidity range from 0.2 to 5.8, we extrapolated
the rapidity distribution. For the O+Au data the extrapolation was based 
on the observed scaling behaviour of 60 GeV/nucleon O+Au data (see section 
5.1.2). The number of net protons in S+Ag collisions at $3.0 \leq
y \leq 5.8$ was estimated to be the average of S+S and S+Au data.  
  
The charge-excess method, corrected for the $K^+/K^-$ asymmetry and strange
hyperon decays, leads to the number of {\it observed} net protons.
In order to estimate the total number of net baryons one needs in addition
an estimate of the total number of net hyperons.
Table \ref{hyperons} shows the measured yields of net lambdas
$(\Lambda - \overline{\Lambda})$ measured in the same experiment 
\cite{Bam89,Bam90,Bae93}.
We derived the corresponding correction for the total number of net hyperons
from the $(\Lambda - \overline{\Lambda})$ distributions 
using Wr\'{o}blewski's
empirical rule \cite{Wro85}
$N_{\!Y\!} \, = 1.6 \cdot N_{\!\Lambda\!}$.
 
Multiplying the number of net protons by 2 (in the case of
p-induced reactions only in the target hemisphere) and adding 
the corrected number of hyperons gives an estimate of the total
number of net baryons. These estimates are
shown in Table~\ref{model}. This procedure is valid for 
collisions of {\it isoscalar} nuclei (equal number of neutrons and protons) 
and only a crude approximation
in all other cases. Only in the case of central collisions of
{\it isoscalar} nuclei the number of net baryons equates the
number of participant nucleons. 
Comparing the measured number of net baryons with the number of wounded
nucleons using the standard geometrical approach \cite{Kreso2} one finds a 
discrepancy between
these two numbers in asymmetric systems. This can be explained
by target spectators misidentified as participants or additional participants
due to cascading effects in spectator matter.  

\begin{table}[H]
\caption{Number of observed net protons ($p - \overline{p}$) in
p-nucleus, central deuteron-nucleus and central nucleus-nucleus
collisions. The values in italics are the result of 
an extrapolation. See text for details.} 
\label{participant}
\begin{center}
\begin{tabular}{|c|c|c|c|}
\hline   
\multicolumn{4}{|c|}{ $p - \overline{p}$} \cr
\hline
 Reaction &
 $ 0.2 \leq y \leq 3.0 $ &
 $ 3.0 \leq y \leq 5.8 $ &
 $ 0.2 \leq y \leq 5.8 $ \cr 
\hline
\hline
  p+S  &  -  &  - &  2.4 $\pm$ 0.2  \cr 
\hline
  p+Au &  -  &  - &  5.9 $\pm$ 0.2 \cr
\hline
  d+Au &  -  &  - & 10.7 $\pm$ 0.2  \cr
\hline
  O+Au & 42 $\pm$ 5    &  - &  {\it 47 $\pm$ 5 }  \cr 
\hline
  S+S  & 10.6 $\pm$ 0.8 & 10.6  $\pm$ 0.8 &  21.2 $\pm$ 1.3   \cr
\hline
  S+Ag & 32 $\pm$ 2    &  - &  {\it 43 $\pm$ 3 }  \cr
\hline
  S+Au &  -  &  11.4 $\pm$ 0.5 & - \cr
\hline
\end{tabular}
\end{center}
\end{table}

\begin{table}[H]
\caption{Number of observed net lambdas (${\Lambda-\overline{\Lambda}}$) in
p-nucleus and central nucleus-nucleus
collisions. The values in italics are the result of an extrapolation. 
See text for details.} 
\label{hyperons}
\begin{center}
\begin{tabular}{|c|c|c|c|}
\hline   
\multicolumn{4}{|c|}{ $\Lambda - \overline{\Lambda}$} \cr
\hline
 Reaction &
 $ 1.0 \leq y \leq 5.0 $ &
 $ 1.4 \leq y \leq 4.4 $ &
 $ 0.2 \leq y \leq 5.8 $ \cr 
\hline
\hline
  p+S  &  0.19 $\pm$ 0.02  &  - &  {\it 0.23 $\pm$ 0.03}  \cr 
\hline
  p+Au &  -  &  0.20 $\pm$ 0.02 &  {\it 0.28 $\pm$ 0.03} \cr
\hline
\hline   
  &
 $ 0.5 \leq y \leq 3.0 $ &
 $ 3.0 \leq y \leq 5.5 $ &
 $ 0.2 \leq y \leq 5.8 $ \cr 
\hline
\hline
  S+S  & 3.6 $\pm$ 0.4 & 3.6  $\pm$ 0.4 & {\it 7.2 $\pm$ 0.6}   \cr
\hline
  S+Ag & 8.5 $\pm$ 0.8 & {\it 3.9 $\pm$ 0.5 } &  {\it 12.6 $\pm$ 1.2 }  \cr
\hline
\hline
  & & $ 3.0 \leq y \leq 5.0 $ & \cr 
\hline
\hline
  S+Au &  -  &  3.9 $\pm$ 0.4 & - \cr
\hline
\end{tabular}
\end{center}
\end{table}


\begin{table}[H]
\caption{Estimated total number of net baryons  $< B - \overline{B}> $ in
p-nucleus and central nucleus-nucleus
collisions compared to the number of wounded nucleons $< W >$ 
(the values for p+S and p+Au marked by ($^*$) are averaged over all 
impact parameters).} 
\label{model}
\begin{center}
\begin{tabular}{|c|c|c|}
\hline   
 Reaction &   $< B - \overline{B}> $  &  $ < W > $  \cr
          & $ 0.2 \leq y \leq 5.8 $ &        calculated    \cr 
\hline
\hline
  p+S  & 4.2 $\pm$ 0.8              &   3.2$^*$   \cr 
\hline
  p+Au &   -                         &   4.7$^*$   \cr
\hline
  d+Au &   -                        &   10  \cr
\hline
  O+Au &   -                        &   68  \cr
\hline
  S+S  & 54 $\pm$ 3                 &   53  \cr
\hline
  S+Ag & 105 $\pm$ 12               &   92  \cr
\hline
  S+Au &    -                       &  113  \cr
\hline
\end{tabular}
\end{center}
\end{table}

\newpage

\subsection{Transverse momentum distributions of net protons}

The invariant cross sections as a function of the transverse momentum 
for net protons are compatible 
with the following parametrization of the transverse momentum
spectra as suggested by thermodynamic models:
\begin{equation}
 \frac {1} {p_T}~ \frac {dn} {dp_T} = C \cdot
m_T \cdot K_1{(\frac {m_T} {T_1}) },
\end{equation}
where $m_T = \sqrt{ p_T^2 + m_0^2}$ is the transverse mass of particles with
rest mass $m_0$, $C$ is a normalization factor and $K_1$ is the modified
Bessel function of order one. 
The parameter $T_1$ is determined by fitting the experimental data and 
can be interpreted in thermal models without transverse flow 
as the thermal freeze-out `temperature'.

\begin{figure}[H]
\centerline{\epsfig{file=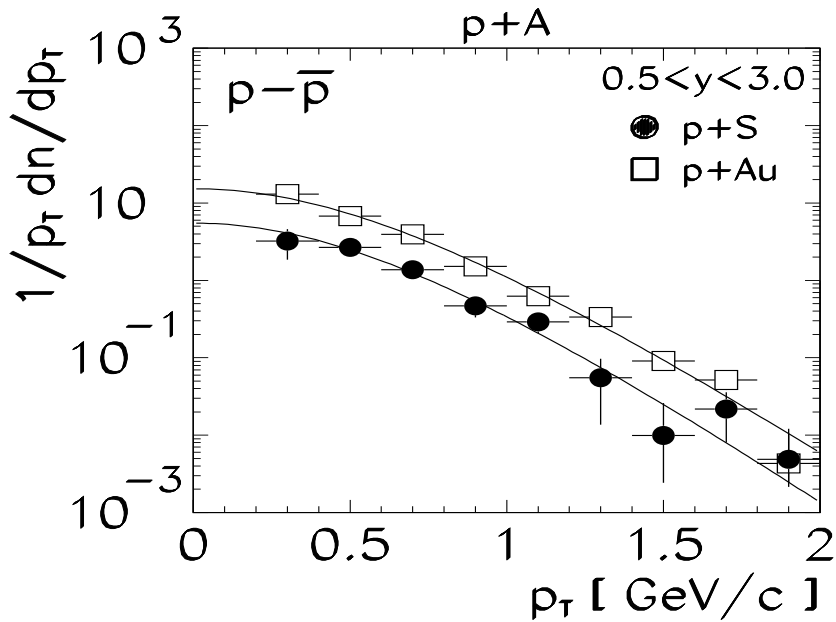,height=9cm,width=12cm}}
\vspace{-1cm}
\caption{Transverse momentum distributions of net protons
for minimum bias p+S and p+Au collisions. The vertical scale is given in (GeV)$^{-2}$.}
\label{22}
\end{figure}

\begin{figure}[H]
\centerline{\epsfig{file=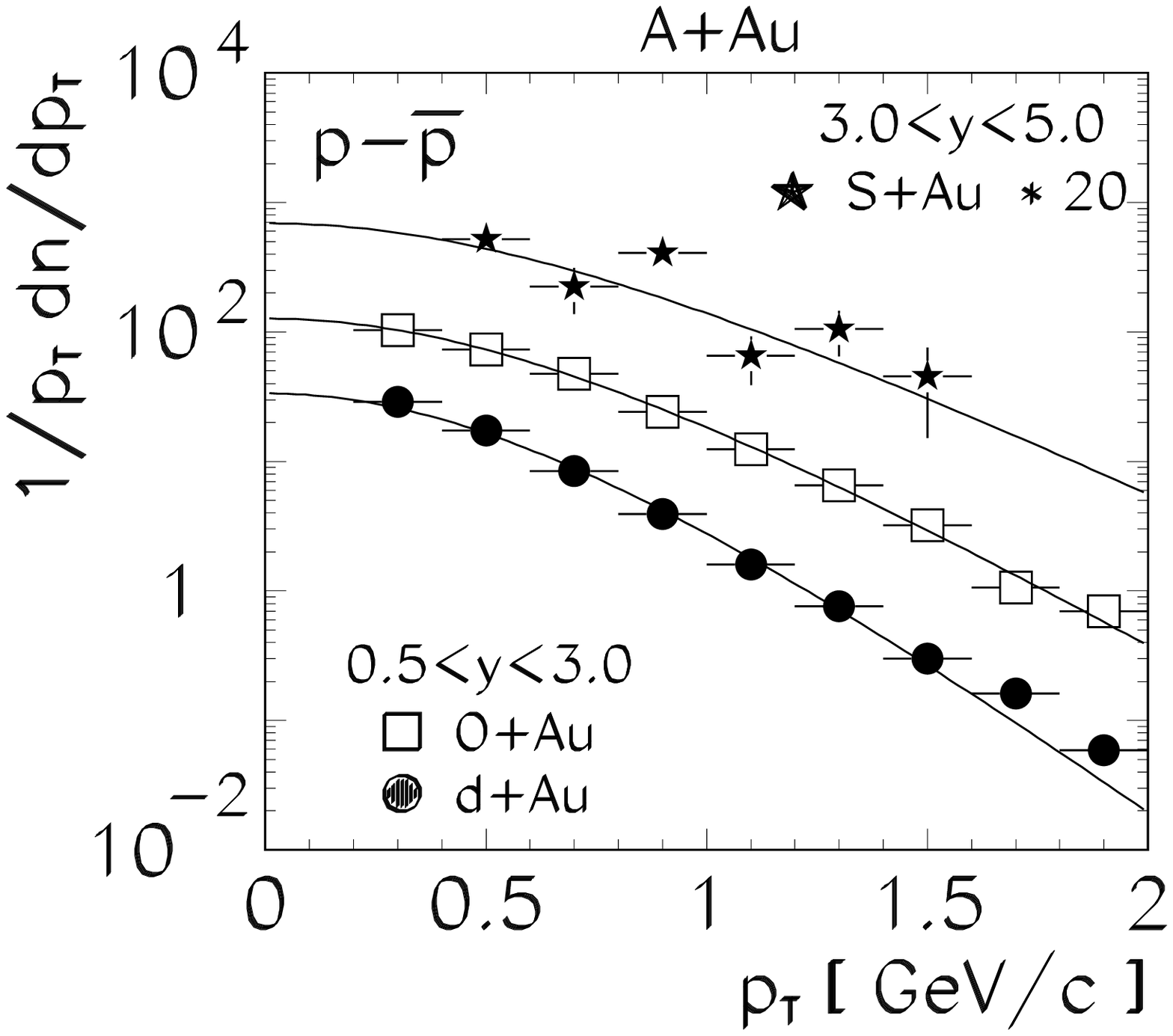,height=9cm,width=12cm}}
\vspace{-1cm}
\caption{Transverse momentum distributions of net protons
for central d+Au, O+Au and S+Au collisions. The vertical scale is given in (GeV)$^{-2}$.}
\label{23}
\end{figure}

\begin{figure}[H]
\centerline{\epsfig{file=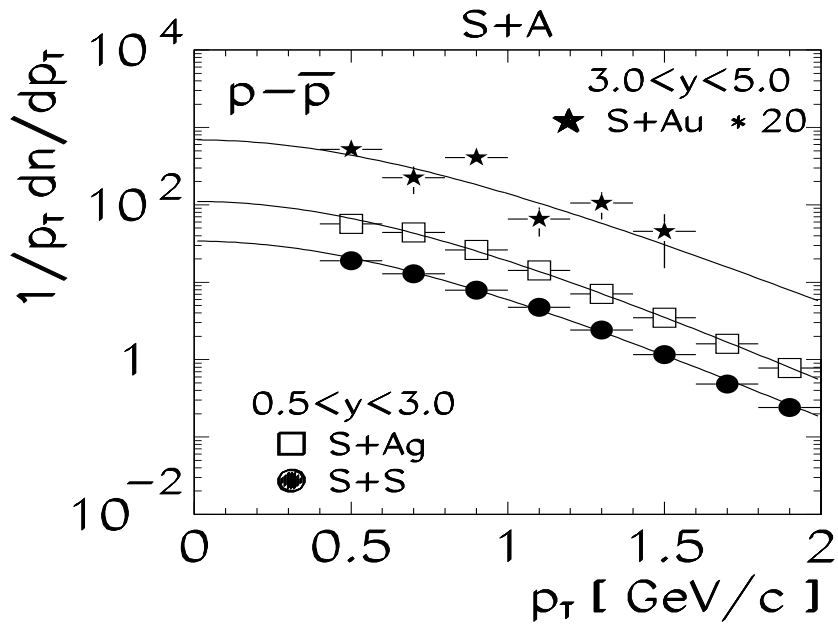,height=9cm,width=12cm}}
\vspace{-1cm}
\caption{Transverse momentum distributions of net protons
for central S+S, S+Ag and S+Au collisions. The vertical scale is given in (GeV)$^{-2}$.}
\label{24}
\end{figure}

%
Figs.~\ref{22}-\ref{24} show transverse momentum
distributions of net protons for various  
collision systems at 200 GeV/nucleon together with fits to Eq.~1. 
The invariant 
spectra of protons as a function of $p_T$ for various systems
are well described by inverse slope parameters (`temperatures') 
of approximately $200$ MeV (Table~\ref{p_temps}). 
The `temperatures' are observed to increase with the mass of the colliding system and are
highest at midrapidity. The `temperatures' range from $T \approx 145$ MeV for
minimum bias p+S to $T \approx 220-240$ MeV 
for central collisions of S+S, S+Ag and S+Au.
Dividing the rapidity range into two bins, 
$ 0.5 \leq y \leq 2.0 $ and $ 2.0 \leq y \leq 3.0 $, and fitting the spectra
separately in these bins results in inverse slope parameters that are about 
$40$ MeV higher at midrapidity as compared to the values at lower rapidity
for all systems. 
%
%
%
%

Fitting a simple exponential 
\begin{equation}
 \frac {1} {p_T}~ \frac {dn} {dp_T} = C \cdot
 \exp{(- \frac {m_T} {T_2}) }
\end{equation}
results in somewhat higher `temperatures' as shown in Table \ref{p_temps}. 
Also shown are our published results for S+S collisions and results from 
another experiment.
The `temperature' reported by NA35 in \cite{Bae94} was about 40 MeV lower. 
Much higher statistics of the recent data allowed a more detailed contamination
correction as described in Section 4. NA44 data \cite{Dodd,Bearden} measured
in a small rapidity interval near midrapidity for less central collisions 
yield somewhat lower inverse slope parameters.   

\begin{table}[H]
\caption{`Temperatures' obtained from thermal fits (Eq.~1) and fits 
according to Eq.~2 to the
transverse momentum distributions of participant protons in
various collisions. The systematic error is about 10 MeV.} 
\label{p_temps}
\begin{center}
\begin{tabular}{|c|c|c|c|c|}
\hline   
 &  \multicolumn{2}{|c|}{$T_1$ [MeV]} &  \multicolumn{2}{|c|}{$T_2$ [MeV]} \cr
 &  \multicolumn{2}{|c|}{(Eq.~1)}     &  \multicolumn{2}{|c|}{(Eq.~2)} \cr
\hline
 Reaction &
 $ 0.5 \leq y \leq 3.0 $ &
   NA35  &
 $ 0.5 \leq y \leq 3.0 $ &
   NA44      \cr
\hline
\hline
  p+S  & 145 $\pm$ 11 &  & 154 $\pm$ 14 & \cr 
\hline
  p+Au & 156 $\pm$ 4  &  & 163 $\pm$ 5  & \cr 
\hline
  d+Au & 161 $\pm$ 2  &  & 172 $\pm$ 5  & \cr 
\hline
  O+Au & 204 $\pm$ 4  &  & 219 $\pm$ 5  & \cr 
\hline
  S+S  & 226 $\pm$ 5  & 180 $\pm$ 14 \cite{Bae94} & 235 $\pm$ 9  
                 & 208 $\pm$ 8 \cite{Bearden} \cr
\hline
  S+Ag & 224 $\pm$ 2  &  & 238 $\pm$ 2  & \cr
\hline
\hline
      & $ 3.0 \leq y \leq 5.0 $ &  & $ 3.0 \leq y \leq 5.0 $  & \cr
\hline
\hline
  S+Au & 244 $\pm$ 43 &  & 276 $\pm$ 48 & 242 $\pm$ 3 \cite{Dodd} \cr
\hline
\end{tabular}
\end{center}
\end{table}

\newpage

\section{Negatively Charged Hadron Production}


Essentially all of the energy deposited in the reaction volume 
by the participant
nucleons goes into the production of hadrons.
Without extensive particle identification
in this experiment, only distributions of inclusive charged hadrons  
can be measured and separation of various species (such as 
$\pi^+$, K$^+$, $p$) is not possible. 
The positive hadron distributions are more complicated to interpret
than the negative ones, since the contribution of $K^+$ and $p$ 
is a large contamination. 
Thus, we will concentrate on the distributions of negative hadrons
which are primarily $\pi^-$ mesons. The relevant contamination and
acceptance corrections as described in section 4 were applied.

\subsection{Rapidity Distributions}

\subsubsection{Proton-Nucleus Collisions}

\begin{figure}[H]
\centerline{\epsfig{file=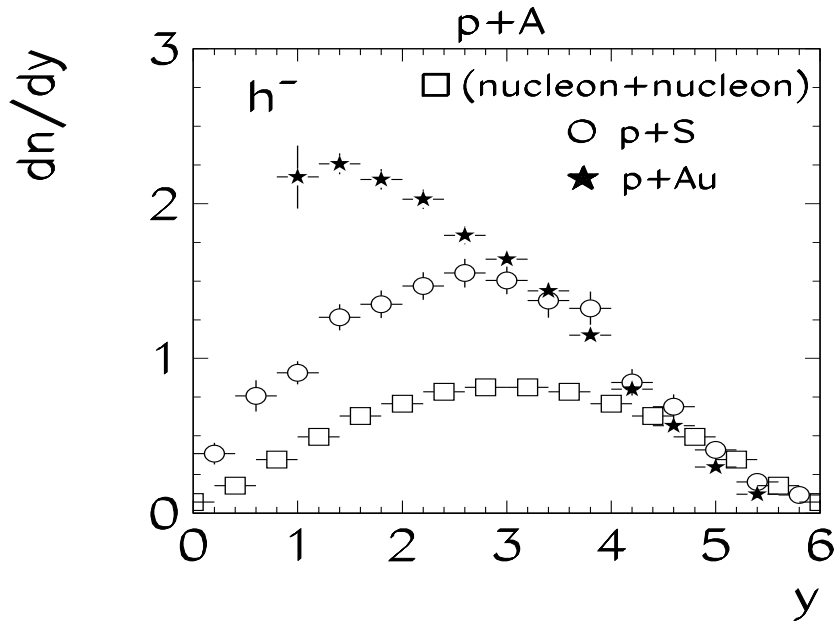,height=9cm,width=12cm}}
\vspace{-1.0cm}
\caption{Rapidity distributions of negatively charged hadrons
produced in minimum bias nucleon--nucleon, p+S and p+Au collisions.}
\label{66}
\end{figure}

\begin{figure}[H]
\centerline{\epsfig{file=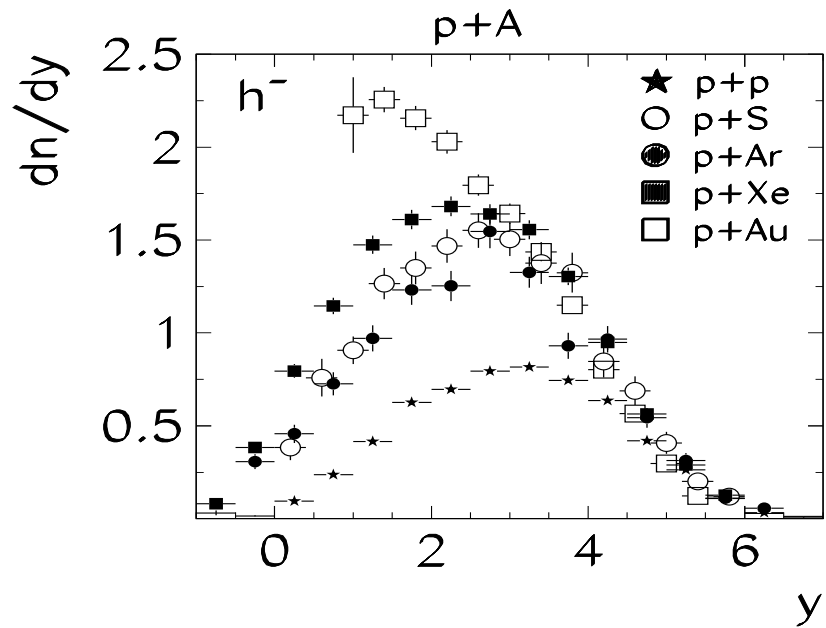,height=9cm,width=12cm}}
\vspace{-1.0cm}
\caption{Rapidity distributions of negatively charged hadrons
produced in minimum bias p+p, p+S, p+Ar, p+Xe and p+Au collisions 
\protect\cite{Demarzo}. Note that the rapidity distribution for p+S
collisions is higher than for p+Ar interactions due to different 
trigger conditions and event selections.}
\label{65}
\end{figure}

The rapidity distributions of negatively charged hadrons produced in
p+S and p+Au interactions are shown in Fig.~\ref{66}. For
comparison nucleon--nucleon reactions are also included. 
The nucleon--nucleon interactions are defined here as 
$0.5 \cdot [ (p+p \rightarrow \pi^+) + (p+p \rightarrow \pi^-) ]$, 
the data are taken from \cite{Kafka} 
(cf. \cite{Adamus} and \cite{Gaz91} and references therein).
The $h^-$ from 
nucleon--nucleon interactions peak at midrapidity. 
The multiplicity increases with target mass, and the maximum of
the distribution shifts to lower rapidities with increasing target mass.
This is indicative of $\pi^-$-production after multiple collisions. 
For $y \ge 4$,  the yield of $h^-$ produced in p+Au collisions
appears to be lower than that for nucleon--nucleon and p+S collisions.  
Fig.~\ref{65} shows a comparison to minimum bias p+p, p+Ar and p+Xe data
from experiment NA5 at the CERN-SPS \cite{Demarzo}.     

\newpage

\subsubsection{Nucleus-Gold Collisions}

\begin{figure}[H]
\centerline{\epsfig{file=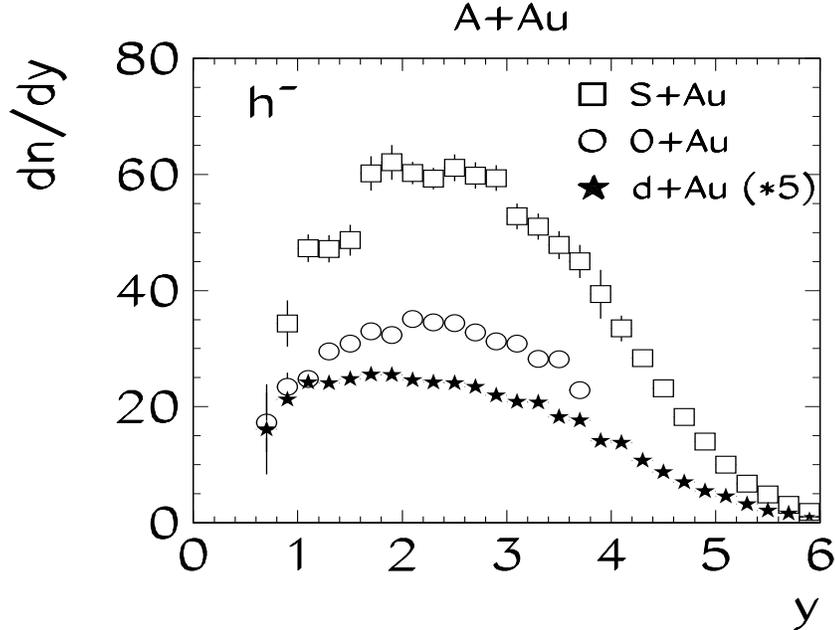,height=9cm,width=12cm}}
\vspace{-1.5cm}
\caption{Rapidity distributions of negatively charged hadrons
produced in central d+Au, O+Au and S+Au collisions.}
\label{67}
\end{figure}

The rapidity distributions of negatively charged hadrons produced in
central d+Au, O+Au and S+Au collisions 
are shown in Fig.~\ref{67}. 
The distributions exhibit a peak at $y$ below $y_{cm} = 3$. 
The multiplicity increases with projectile mass, and the maximum of
the distribution moves toward midrapidity with increasing projectile mass.
%


\subsubsection{Sulphur-Nucleus Collisions}

\begin{figure}[H]
\centerline{\epsfig{file=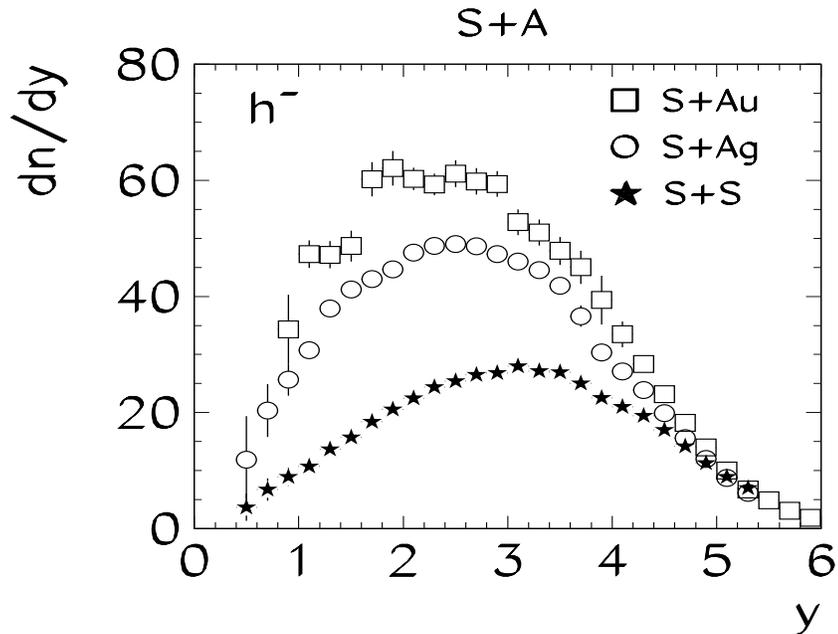,height=9cm,width=12cm}}
\vspace{-1.5cm}
\caption{Rapidity distributions of negatively charged hadrons
produced in central S--nucleus collisions.}
\label{68}
\end{figure}

\begin{figure}[H]
\centerline{\epsfig{file=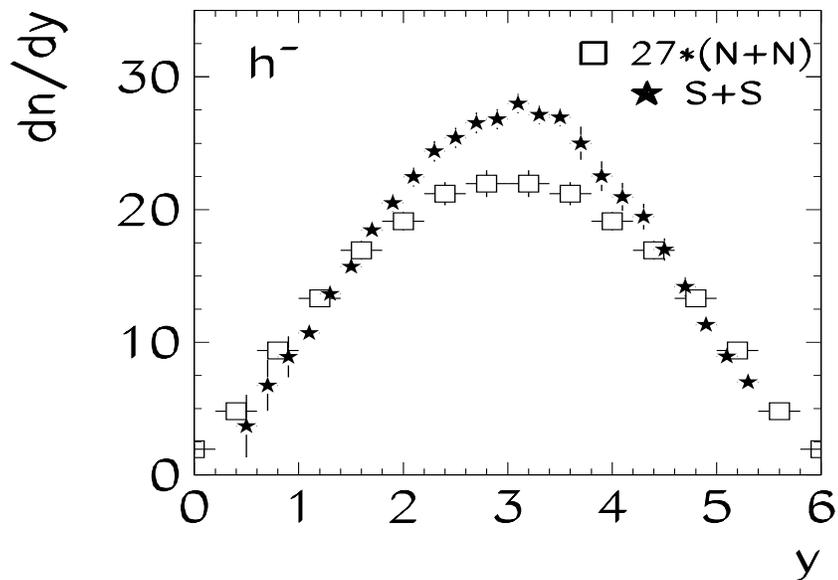,height=9cm,width=12cm}}
\vspace{-1.5cm}
\caption{Rapidity distributions of negatively charged hadrons
produced in central S--S collisions and in minimum bias nucleon--nucleon
interactions (scaled by the ratio of participant nucleons).}
\label{69}
\end{figure}

To investigate the target mass dependence for S-induced reactions, 
rapidity distributions of negatively charged hadrons produced in
central S+S, S+Ag and S+Au collisions 
are compared in Fig.~\ref{68}. 
The $h^-$ distribution peaks at midrapidity for the symmetric 
S+S interactions. 
The multiplicity increases with the target mass, and the maximum of
the distribution moves to lower rapidities with increasing target mass.
Fig.~\ref{69} shows a comparison of the rapidity distributions of 
negatively charged hadrons
produced in central S+S collisions and in minimum bias nucleon--nucleon
interactions. The latter was scaled by the ratio of participant nucleons. 
The width of the distribution for S+S collisions is slightly narrower
than the one for the nucleon--nucleon interactions.
The mean negative hadron multiplicity per participant is higher in
central S+S collisions than in N+N interactions by about $20\%$.

%


\subsubsection{Extrapolation to $4\pi$}

Since the rapidity distributions were measured over a wide kinematic
range and cover almost all rapidities between target and beam rapidity,
the spectra can be extrapolated to full phase space in order to obtain
average total multiplicities. The results are compiled in Table~\ref{mult}. 
The extrapolation to target and
beam rapidity was based on a gaussian fitted locally to the adjacent part
of the rapidity distribution. The multiplicities increase with system mass. 
For comparison average total multiplicities reported from different experiments
are also shown. For a review of multiplicities of negatively charged 
hadrons produced in nuclear collisions cf. \cite{Marek}.

\begin{figure}[H]
\centerline{\epsfig{file=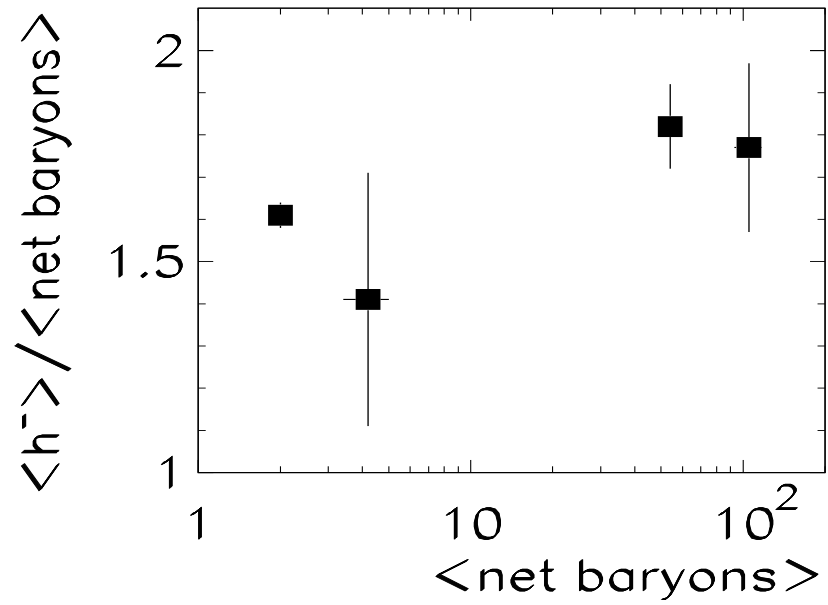,height=9cm,width=12cm}}
\vspace{-1.0cm}
\caption{Average total multiplicity of
negatively charged hadrons per net baryon 
produced in
minimum bias nucleon--nucleon, p+S, central S+S and S+Ag 
collisions as a function of the number of net baryons.}
\label{hmin_per_netb}
\end{figure}

The average multiplicity of negatively charged hadrons increases with 
the number of net protons.
For not too asymmetric systems the ratio of the multiplicity of
negatively charged hadrons per net (participant) baryon is shown
in Fig.~\ref{hmin_per_netb}. This ratio is smaller in
nucleon--nucleon and p+S interactions than in central S+S and S+Ag collisions. 

\newpage
\begin{table}[H]
\caption{Average multiplicity (in acceptance and total) of negatively charged 
hadrons produced in minimum bias nucleon--nucleon (N+N),
minimum bias p--nucleus, deuteron--nucleus and central nucleus--nucleus
collisions. The first column lists the reaction, the second column is the 
measured rapidity range and corresponding average multiplicity and the third column
is the extrapolation to $4\pi$ acceptance.} 
\label{mult}
\begin{center}
\begin{tabular}{|c|c|c|}
\hline   
\multicolumn{3}{|c|}{ $< h^- >$} \cr
\hline
 Reaction & $ 0.0 \leq y \leq 6.0 $ & $4\pi$ \cr
\hline
\hline
  p+p \protect\cite{Gaz91} &  -   &  2.85 $\pm$ 0.03  \cr 
\hline
  N+N \protect\cite{Gaz91} &  -   &  3.22 $\pm$ 0.06  \cr 
\hline
\hline
  p+Mg \protect\cite{Brick} &   -  &  4.9 $\pm$ 0.4  \cr 
\hline
\hline
  p+S \protect\cite{Bam90}  &  -   &  5.0 $\pm$ 0.2  \cr 
\hline
  p+S  &  5.7  $\pm$ 0.1 &  5.9 $\pm$ 0.2  \cr 
\hline
\hline
  p+Ar \protect\cite{Demarzo}  &  -  &  5.39 $\pm$ 0.17  \cr 
\hline
\hline
  p+Ag \protect\cite{Brick} &   -  &  6.2 $\pm$ 0.2  \cr 
\hline
\hline
  p+Xe \protect\cite{Demarzo}  &  -  &  6.84 $\pm$ 0.13  \cr 
\hline
\hline
          & $ 0.8 \leq y \leq 6.0 $ & $4\pi$ \cr
\hline
\hline
  p+Au &  6.6 $\pm$ 0.1  & 9.9 $\pm$ 2 \cr
\hline
  p+Au \protect\cite{Brick} &   -  &  7.0 $\pm$ 0.4  \cr 
\hline
  p+Au \protect\cite{Bae91a} &   -  &  7.3 $\pm$ 0.3  \cr 
\hline
  `central' &     &   \cr 
  p+Au \protect\cite{Bam89} &   -  &  9.6 $\pm$ 0.2  \cr 
\hline
\hline
          & $ 0.6 \leq y \leq 6.0 $ & $4\pi$ \cr
\hline
  d+Au & 17.2 $\pm$ 0.3  & 23 $\pm$ 4  \cr
\hline
\hline
          & $ 0.6 \leq y \leq 3.8 $ & $4\pi$ \cr
\hline
\hline
  O+Au & 94 $\pm$ 2      & 137 $\pm$ 9 \cr 
\hline
  O+Au \protect\cite{Bae91a}&  -    & 124.0 $\pm$ 1.9 \cr 
\hline
\hline
          & $ 0.4 \leq y \leq 5.4 $ & $4\pi$ \cr
\hline
\hline
  S+S  &  90 $\pm$ 1     & 98 $\pm$ 3  \cr
\hline
  S+S \protect\cite{Bae94}&  -   & 94 $\pm$ 5 \cr 
\hline
  S+Ag & 162 $\pm$ 2     & 186 $\pm$ 11 \cr
\hline
\hline
          & $ 0.8 \leq y \leq 6.0 $ & $4\pi$ \cr
\hline
  S+Au & 201 $\pm$ 2     & 225 $\pm$ 12 \cr
\hline
\end{tabular}
\end{center}
\end{table}

%
%
%

\newpage

\subsection{Transverse Momentum Spectra}


It is important to study the transverse degrees of freedom for produced
hadrons in order to learn more about the degree of thermalization in heavy ion
collisions and possible effects of collective flow on the spectra. 
Thermal model fits of the invariant cross 
sections (see Eqs.~1 and 2)
as a function of $p_T$ provide information on the 
production mechanisms for hadrons.  
It is known that pions, the majority of the negatively charged
hadrons, are to a large part daughters from the decays of
heavier resonances. In addition, possible transverse 
collective flow could further modify the spectra. It is thus
not too surprising that 
the $p_T$ spectra 
of negatively charged hadrons are, in general,  
incompatible with thermal parametrizations  
over the entire $p_T$-range with the exception of p+S interactions
\cite{Alp75,Gue76b,Cro75,Ake90,Bogg}.
Since the $p_T$ distributions are found not to be thermal, 
a systematic analysis
using Eq.~2 fits is excluded. Therefore, the $h^-$ data on 
transverse momentum
distributions will be presented systematically without fits in the
following sections. However, a fit to Eq.~2 was performed for
all systems in the rapidity bin near midrapidity in a limited
$p_T$ range ($0.3 \le p_T \le 1.0$ GeV/c) in order to give a 
rough estimate of `temperatures'.

\subsubsection{Proton--Nucleus collisions}

\begin{figure}[H]
\centerline{\epsfig{file=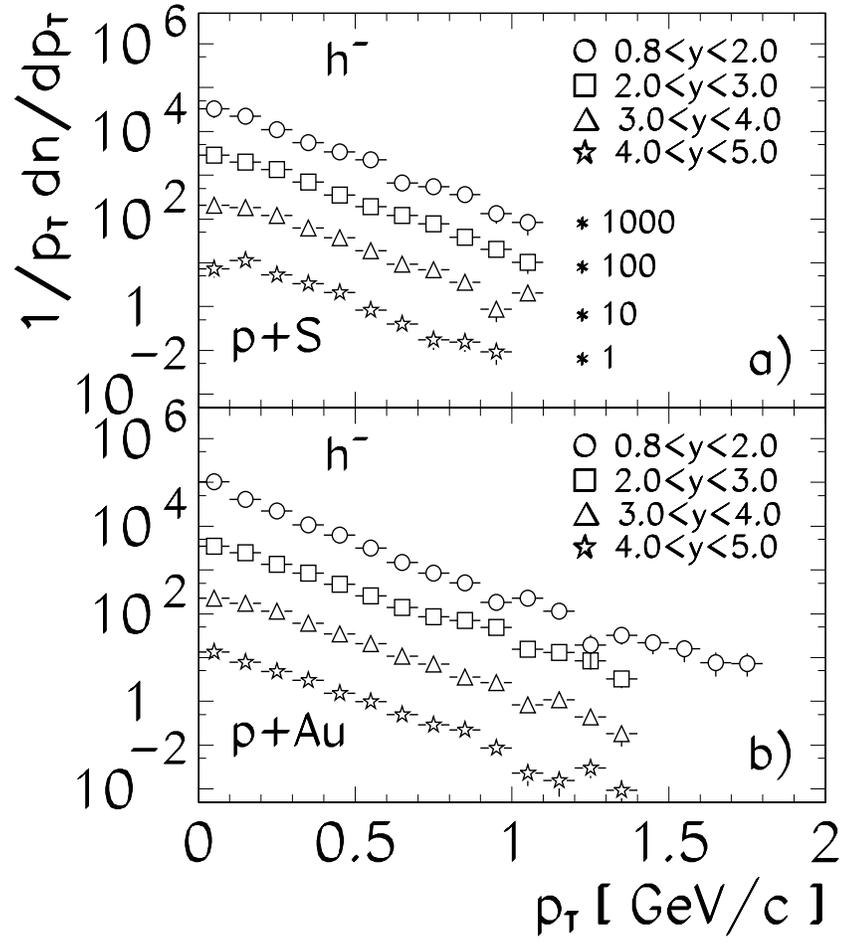,height=14cm,width=12cm}}
\vspace{-1.0cm}
\caption{Transverse momentum distributions of negatively charged
hadrons produced in p+S (a) and p+Au (b) interactions 
for the various rapidity intervals shown. The vertical scale is given in (GeV)$^{-2}$.}
\label{241}
\end{figure}

The transverse momentum spectra of negatively charged hadrons produced
in minimum bias p+S and p+Au collisions are shown in
Figs.~\ref{241}~a,b for four different rapidity intervals:
$0.8 < y < 2.0$, $2.0 < y < 3.0$, $3.0 < y < 4.0$, $4.0 < y < 5.0$. The 
corresponding 
spectra have been multiplied by the factors of 1000, 100, 10, 
and 1, respectively, for presentation on the same plot.
\vspace{-0.5cm}
%
%
%
%
%

\newpage

\subsubsection{Nucleus-Nucleus Collisions}

\begin{figure}[H]
\vspace{-1.6cm}
\centerline{\epsfig{file=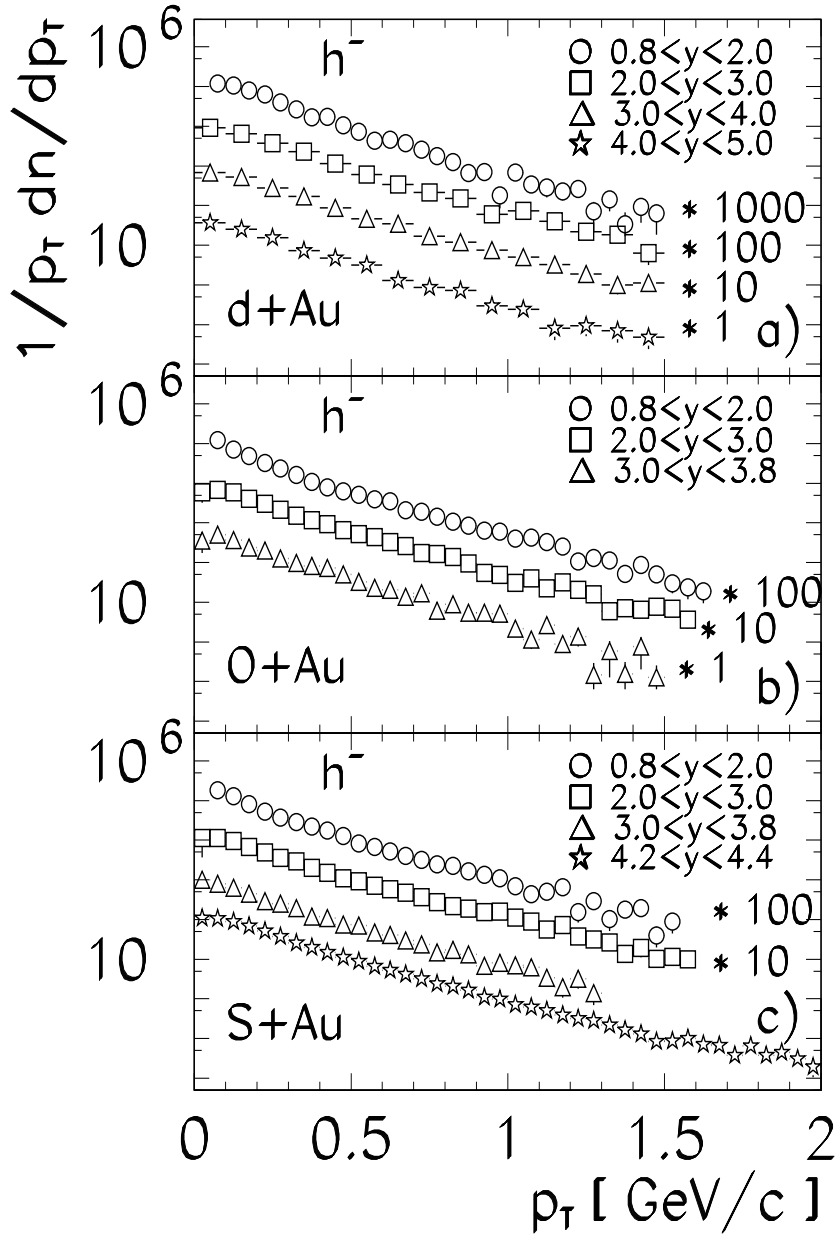,height=18cm,width=12cm}}
\vspace{-1.0cm}
\caption{Transverse momentum distribution of negatively charged
hadrons produced in central d+Au (a), central O+Au (b) and 
central S+Au (c) collisions at different rapidities. 
The vertical scale is given in (GeV)$^{-2}$.}
\label{242}
\end{figure}

The transverse momentum spectra of negatively charged hadrons produced
in central d+Au interactions and central O+Au and S+Au collisions 
are shown in
Figs.~\ref{242}~a-c for various rapidity intervals.
The corresponding 
spectra were multiplied by factors of 1000, 100, 10, and 1, 
respectively, in the case of d+Au collisions. For the heavy ion collisions
the first two spectra (rapidities below midrapidity) were multiplied 
by factors of 100 and 10.
The distributions especially at low rapidities ($0.8 < y < 2.0$)
for central O+Au and S+Au collisions clearly exhibit a 
deviation from a single exponential in $p_T$ with an enhancement at low 
transverse momenta in the backward rapidity region. 

%
%
%

\begin{figure}[H]
\centerline{\epsfig{file=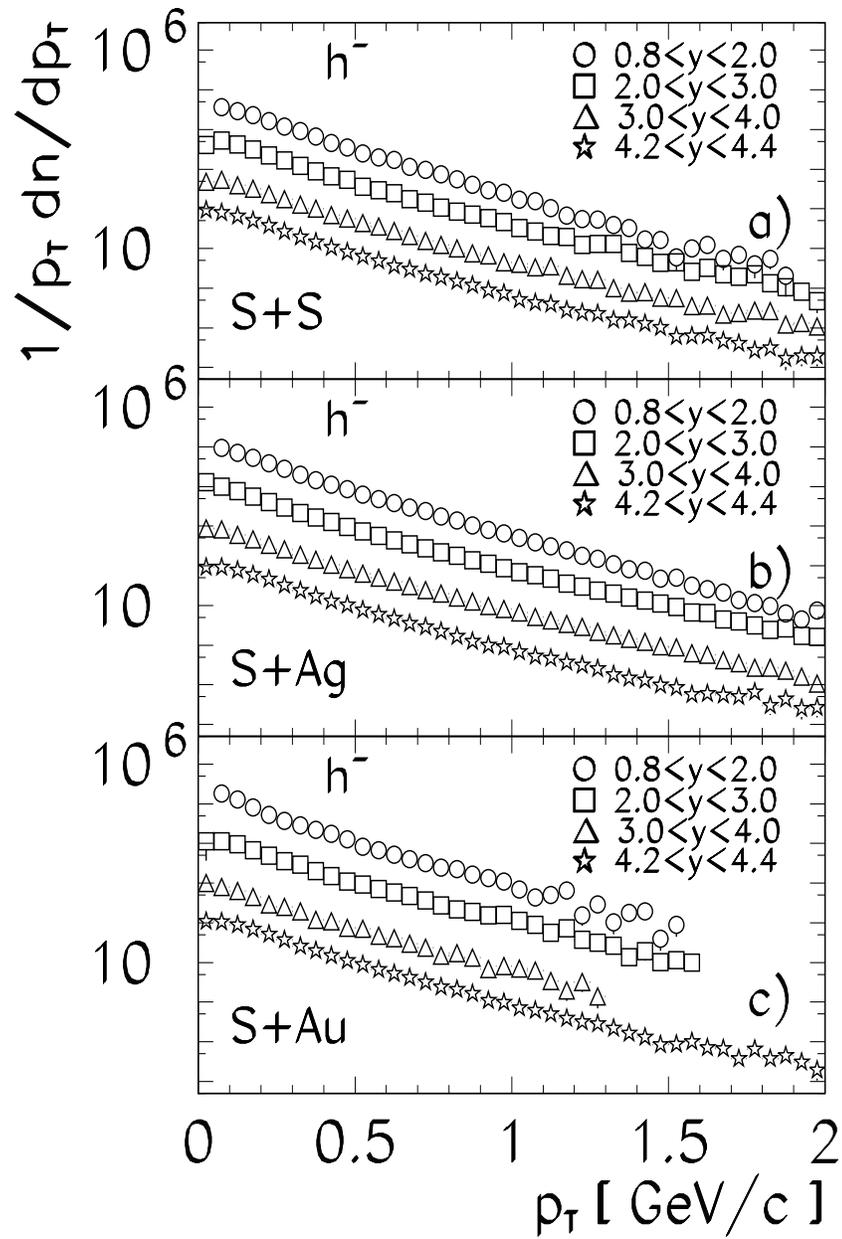,height=18cm,width=12cm}}
\vspace{-1.0cm}
\caption{Transverse momentum distribution of negatively charged
hadrons produced in central S+S (a), S+Ag (b) and S+Au (c) collisions 
at different
rapidities. The vertical scale is given in (GeV)$^{-2}$.}
\label{244}
\end{figure}

The transverse momentum spectra of negatively charged hadrons produced
in central S+S, S+Ag and S+Au collisions are shown in
Figs.~\ref{244}~a-c for various rapidity intervals.

%
%
%
%
%
%


\subsection{Mean transverse momentum}

\begin{figure}[H]
\centerline{\epsfig{file=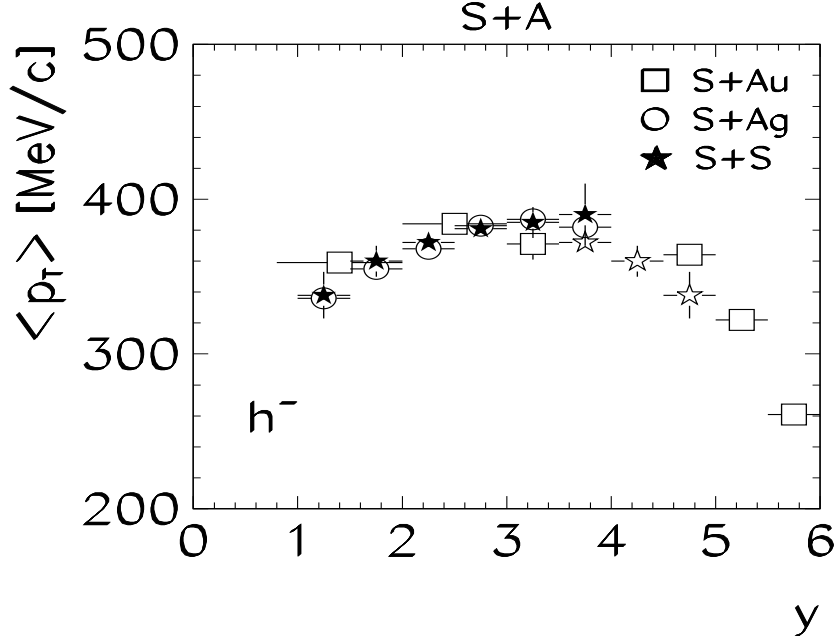,height=9cm,width=12cm}}
\vspace{-1.0cm}
\caption{Mean transverse momentum of negatively charged
hadrons produced in central S+S, S+Ag and S+Au collisions as a function of
rapidity. The open stars are S+S data reflected at $y = 3$.}
\label{sa_meanpt}
\end{figure}

The invariant cross sections of negatively charged hadrons cannot be
fitted over the entire $p_T$ range with a single exponential in $p_T$ or $m_T$.
Thus, the mean transverse momentum is calculated for various systems and 
rapidity intervals, and the
results are presented in Table~\ref{mean_pt}. As an example, 
the rapidity dependence of $\langle p_T \rangle$ in central S+A collisions 
is shown in Fig.~\ref{sa_meanpt}.
The mean transverse momentum reaches a maximum at midrapidity
and exhibits little dependence on rapidity in the vicinity of $y = 3$.

\begin{figure}[H]
\centerline{\epsfig{file=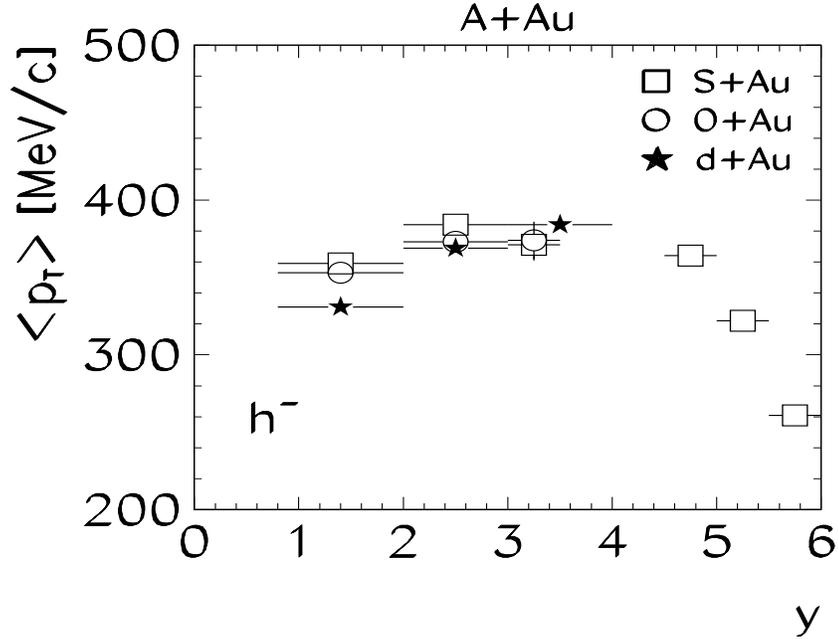,height=9cm,width=12cm}}
\vspace{-1.0cm}
\caption{Mean transverse momentum of negatively charged
hadrons produced in d+Au, central O+Au and S+Au collisions as a function of
rapidity.}
\label{aau_meanpt}
\end{figure}

For $y > 4.5$,  $\langle p_T \rangle$ decreases rapidly.
For the gold target (Fig.~\ref{aau_meanpt}) there is a slight 
increase of 
$\langle p_T \rangle$ in the target hemisphere with the projectile mass.
For p+p interactions the $\langle p_T \rangle$ is
340 $\pm$ 7 MeV/c for $\pi^-$ \cite{Ros75,Kafka,Whit} and 366 $\pm$ 2 MeV/c for
all charged particles in $1.5 < y < 4.5$ \cite{Demar84}.  

\vspace{-0.5cm}
%
%
%
\begin{table}[H]
\caption{Mean transverse momenta of negatively charged hadrons 
produced in
p--nucleus, d--nucleus and central nucleus--nucleus
collisions for various rapidity intervals. The systematic error is less than
10 MeV/c.} 
\label{mean_pt}
\begin{center}
\begin{tabular}{|c|c|c|c|}
\hline 
\multicolumn{4}{|c|}{ $ < p_T > [MeV/c]$} \cr
\hline  
 Reaction &
 $ 0.8 \leq y \leq 2.0 $ &
 $ 2.0 \leq y \leq 3.0 $ & 
 $ 3.0 \leq y \leq 4.0 $ \cr
\hline
\hline
  p+S  & 325 $\pm$ 6 & 363 $\pm$ 8 & 335 $\pm$ 8 \cr 
\hline
  p+Au & 311 $\pm$ 5 & 377 $\pm$ 6 & 372 $\pm$ 7 \cr
\hline
  d+Au & 331 $\pm$ 4 & 369 $\pm$ 5 & 384 $\pm$ 5 \cr
\hline
\hline
 &
 $ 0.8 \leq y \leq 2.0 $ &
 $ 2.0 \leq y \leq 3.0 $ &
 $ 3.0 \leq y \leq 3.5 $ \cr
\hline
\hline
  O+Au & 353 $\pm$ 5 & 373 $\pm$ 4 & 374 $\pm$ 12 \cr 
\hline
  S+S  & 349 $\pm$ 9 & 377 $\pm$ 4 & 388 $\pm$ 11 \cr
\hline
  S+Ag & 346 $\pm$ 2 & 376 $\pm$ 1 & 385 $\pm$ 1  \cr
\hline
  S+Au & 359 $\pm$ 6 & 384 $\pm$ 5 & 371 $\pm$ 10  \cr
\hline
\hline
 &
 $ 4.5 \leq y \leq 5.0 $ &
 $ 5.0 \leq y \leq 5.5 $ &
 $ 5.5 \leq y \leq 6.0 $ \cr
\hline
\hline
  S+Au & 364 $\pm$ 2 & 322 $\pm$ 3 & 261 $\pm$ 4 \cr
\hline
\end{tabular}
\end{center}
\end{table}

\subsection{Inverse slope parameters}

\begin{figure}[H]
\centerline{\epsfig{file=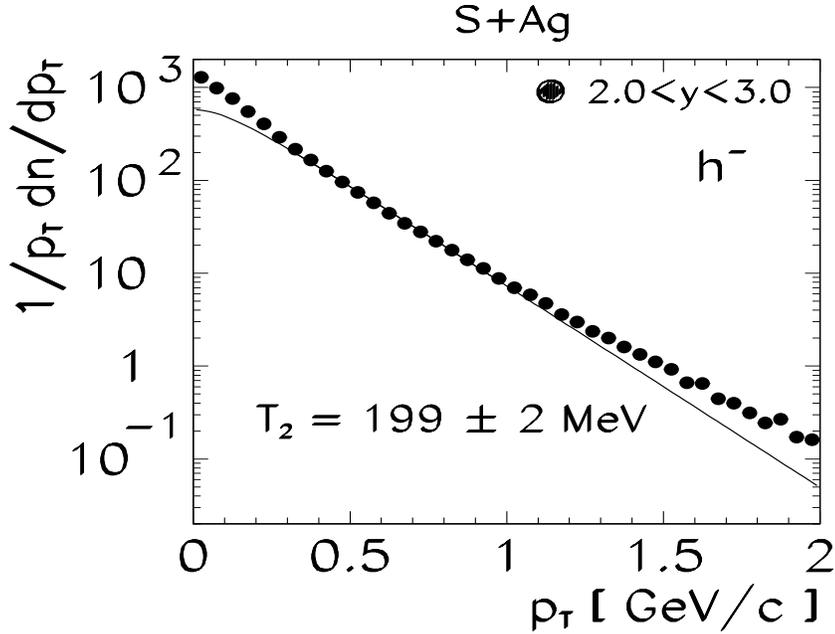,height=9cm,width=12cm}}
\vspace{-1.0cm}
\caption{Transverse momentum distribution of negatively charged
hadrons produced in central S+Ag collisions. The solid line is a fit
of Eq.~2 to the data in the interval $0.3 < p_T < 1.0$ GeV/c. 
The vertical scale is given in (GeV)$^{-2}$.}
\label{mt-fit}
\end{figure}

To facilitate comparison with other published results (see 
Table~\ref{slopes_comp}) fits to
Eq.~2 were performed in the interval $0.3 \le p_T \le 1.0$ GeV/c.
The resulting inverse slope parameters $T$ are listed in Table~\ref{slopes}.
A typical result of such a fit is shown in Fig.~\ref{mt-fit} which indicates
an overall discrepancy between the concavely shaped data and the rather
straight to convex curve corresponding to Eq.~2. The fit is thus
carried out in the restricted range only, and this has always to be kept
in mind in comparing to other results concerning $h^-$ and $\pi^-$.
Below 250 MeV/c the data points are all above the fit. This enhancement
({\it low $p_T$ enhancement}) is most pronounced at low rapidities
in collisions with heavy targets \cite{Roe2}. 
Fitting p+p data in a similar way and averaging the results from 
different experiments yields an inverse slope parameter 
at midrapidity of 142 $\pm$ 10 MeV \cite{Bearden,Adamus,Alp75,Gue76b,Cho}. 

\begin{table}[H]
\caption{Inverse slope parameters $T_2$ (Eq.~2, fitted to the data in
the interval $0.3 \le p_T \le 1.0$ GeV/c) of
negatively charged hadrons produced in
p-nucleus, d-nucleus and central nucleus-nucleus
collisions for various rapidity intervals.} 
\label{slopes}
\begin{center}
\begin{tabular}{|c|c|c|c|c|}
\hline 
\multicolumn{5}{|c|}{ $ T_2 $  [MeV]} \cr
\hline 
 Reaction &
 $ 0.8 \leq y \leq 2.0 $ &
 $ 2.0 \leq y \leq 3.0 $ & 
 $ 3.0 \leq y \leq 4.0 $ &
 $ 4.0 \leq y \leq 5.0 $ \cr
\hline
\hline
  p+S  & 159 $\pm$ 13 & 169 $\pm$ 13 & 154 $\pm$ 10 & 142 $\pm$ 12 \cr 
\hline
  p+Au & 150 $\pm$ 8 & 183 $\pm$ 10 & 177 $\pm$ 9 & 171 $\pm$ 10 \cr
\hline
  d+Au & 151 $\pm$ 9 & 165 $\pm$ 10 & 181 $\pm$ 12 & 183 $\pm$ 11 \cr
\hline
\hline
 &
 $ 0.8 \leq y \leq 2.0 $ &
 $ 2.0 \leq y \leq 3.0 $ &
 $ 3.0 \leq y \leq 3.8 $ &
 $ 4.2 \leq y \leq 4.4 $ \cr
\hline
\hline
  O+Au & 189 $\pm$ 5 & 188 $\pm$ 8  & 175 $\pm$ 10  & - \cr 
\hline
  S+S  & 175 $\pm$ 4 & 191 $\pm$ 7  & 191 $\pm$ 6  & 186 $\pm$ 5 \cr
\hline
  S+Ag & 184 $\pm$ 2 & 199 $\pm$ 2  & 201 $\pm$ 4  & 185 $\pm$ 5 \cr
\hline
  S+Au & 182 $\pm$ 7 & 200 $\pm$ 10 & 190 $\pm$ 10 & 187 $\pm$ 5 \cr
\hline
\end{tabular}
\end{center}
\end{table}

\newpage

\begin{table}[H]
\caption{Inverse slope parameters $T_2$ (Eq.~2, fitted to the data in
the interval $0.3 \le p_T \le 1.0$ GeV/c) of
negatively charged hadrons in  
comparison with other experiments.} 
\label{slopes_comp}
\begin{center}
\begin{tabular}{|c|c|c|}
\hline 
\multicolumn{3}{|c|}{ $ T_2 $ [MeV]} \cr
\hline        & this paper &  NA34 \protect\cite{Ake90} \cr 
              &  $h^-$     &  $h^-$ \cr 
 Reaction &
 $ 0.8 \leq y \leq 2.0 $ &
 $ 1.0 \leq y \leq 1.9 $ \cr 
\hline
\hline
  p+Au/W & 150 $\pm$ 8 &  161 $\pm$ 5 \cr
\hline
  O+Au/W & 189 $\pm$ 5 &  167 $\pm$ 5 \cr 
\hline
  S+Au/W & 182 $\pm$ 7 &  165 $\pm$ 5 \cr
\hline
\hline        & this paper &  NA44 \protect\cite{Dodd,Bearden} \cr 
              & $h^-$      &  $\pi^-$ \cr 
              &            &  $0.25 < m_T-m_0 < 1 $ GeV/c$^2$ \cr 
 Reaction &
 $ 3.0 \leq y \leq 3.8 $ &
 $ 3.0 \leq y \leq 4.1 $ \cr 
\hline
\hline
  S+S     & 191 $\pm$ 6  &  148 $\pm$ 4 \cr
\hline
  S+Au/Pb & 190 $\pm$ 10 &  156 $\pm$ 8 \cr
\hline
\end{tabular}
\end{center}
\end{table}

\begin{figure}[H]
\centerline{\epsfig{file=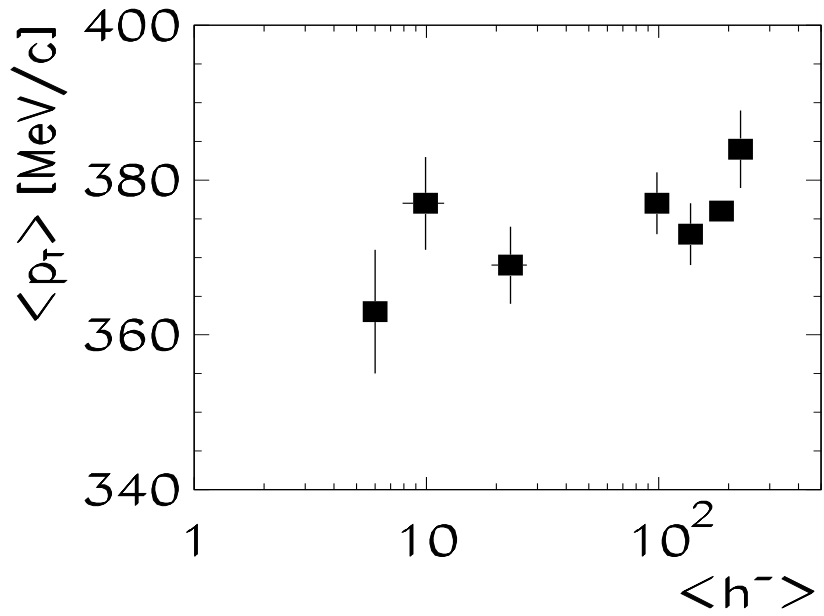,height=9cm,width=12cm}}
\vspace{-1.0cm}
\caption{Average transverse momentum in $2 < y < 3$ of 
negatively charged hadrons produced in
p--nucleus, deuteron--nucleus and nucleus--nucleus
collisions as a function of the average multiplicity.}
\label{meanpt_vs_mult}
\end{figure}

\begin{figure}[H]
\centerline{\epsfig{file=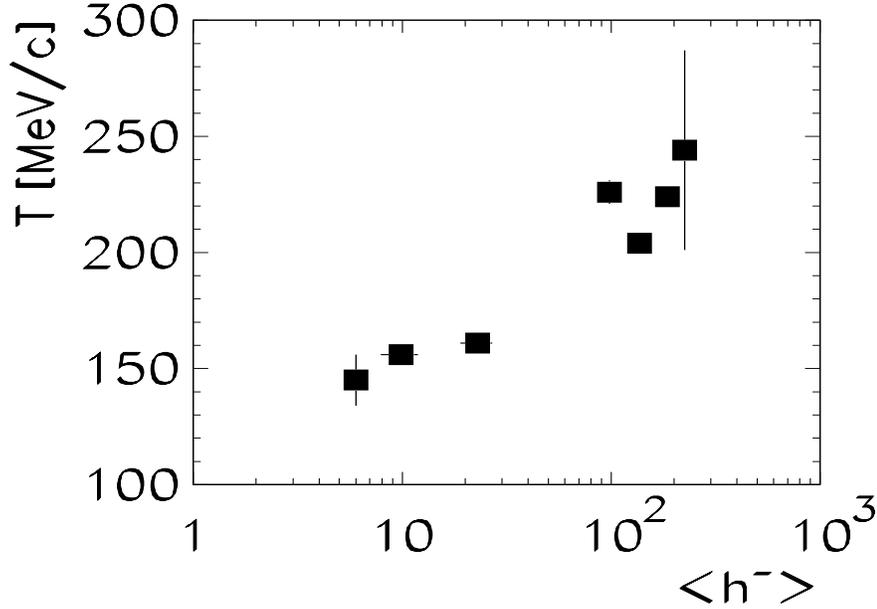,height=9cm,width=12cm}}
\vspace{-1.0cm}
\caption{`Temperature' of net protons for
p--nucleus, deuteron--nucleus and central nucleus--nucleus
collisions as a function of the average multiplicity of negatively 
charged hadrons.}
\label{temp_vs_mult}
\end{figure}

The average transverse momentum of negatively charged hadrons 
exhibits only a slight increase with the system size, measured by
the average total multiplicity of the collision, 
as shown in Fig.~\ref{meanpt_vs_mult}, 
whereas the temperature of the net protons increases from
about 150 MeV in p--nucleus to 200 MeV in central O+Au collisions
and about 240 MeV  in central S--nucleus collisions (Fig.~\ref{temp_vs_mult}).   

\section{Summary}


We have presented results of a systematic study,
using the NA35 experiment at CERN,
of rapidity and transverse momentum 
distributions of net protons and negatively charged
hadrons. A large part of this data was
analyzed employing a new automatic analysis
technique developed by NA35. The most significant findings are:\\
1) The shape of the rapidity distributions
of net projectile protons ($p-\overline{p}$)
is similar
for d+Au, central O+Au and central S+Au collisions.
From this we conclude that the stopping of projectile nucleons
is independent of the size of the (light) projectile nucleus
when incident on a heavy target nucleus (Au in this study).
This suggests that there is no significant difference in the stopping
mechanism as inferred from the proton rapidity
for these light incident nuclei.\\
2) The average rapidity loss
$\langle \delta y \rangle$
for projectile protons
which participate in the interaction
is
$\langle \delta y \rangle$
= 1.2 - 1.6 for various projectiles
from protons to sulphur
incident on a sulphur target
and
$\langle \delta y \rangle$
= 2.0 - 2.3 for a gold target.
The $\langle \delta y \rangle$
are found to be
approximately independent of the
projectile mass for projectile masses ranging from protons
to sulphur. Thus, the rapidity loss of projectile participant protons
increases with the nuclear thickness of the target and is approximately
independent of the nuclear thickness of the lighter projectile as might be expected.\\
3) The number of net protons at midrapidity increases
with projectile mass and with target mass. Since the
mean rapidity shift measured for interacting projectile protons
increases with target mass up to
$\langle \delta y \rangle$
= 2.0 - 2.3 for a gold target
and since the rapidity region for 200 GeV/nucleon
incident energy is a total of $y_{proj} - y_{targ} = 6$,
it is expected that the total number of protons at midrapidity
($y = 3$) will increase with both projectile and target mass.
This results in a
significant pile-up of protons at midrapidity
for the heavier systems.\\
4) The transverse momentum distributions for net protons 
($p-\overline{p}$)
can be described by thermal distributions with `temperatures' ranging from
T $\approx$ 140 MeV for p + S near target rapidity to
T $\approx$ 245 MeV for central S + S and S + Au collisions at midrapidity, using Eq.~1.
Such inverse slope parameters are much higher than the 
Hagedorn temperature ($T \approx 150$ MeV) and cannot be considered
to be true thermal freeze-out temperatures. Our results may indicate the presence
of radially ordered transverse flow in central S+A collisions.\\ 
5) The average negative hadron multiplicity increases with 
both the target and projectile masses. 
The mean negative hadron multiplicity per participant is higher in
central S+A collisions than in N+N interactions by about $20\%$.\\
6) The mean transverse momentum of negatively charged hadrons at midrapidity
is approximately independent of the mass of the incident nuclei.
There is a decrease in the mean transverse momentum when
going away from midrapidity. This is observed for all colliding systems of this
study.\\ 
7) The spectra of negatively charged hadrons do not have a simple thermal shape. 
They exhibit deviations (above a thermal spectrum) at low ($p_T <$ 250 MeV/c)
and high ($p_T >$ 1 GeV/c) transverse momenta. These significant
deviations may be due to
transverse collective flow and resonance production and decay as well as to 
the `Cronin effect'.\\
\\



%
%
%
%

%
{\bf Acknowledgement}\\
We are very grateful to our scanning and measuring crews at
Frankfurt, Munich and Warsaw for their careful and devoted work.
This work was supported by the Bundesministerium f\"ur Forschung
und Technologie and by the Leibniz Grant of the Deutsche Forschungsgemeinschaft,
Germany, by the US Department of Energy (DE--AC03--76SF00098), by the Polish
State Committee for Scientific Research (204369101),
by the Polish-German Foundation (565/93) and
by the Commission of European Communities (CI--0250YU(A)).

\end{document}